\documentclass[a4paper,11pt]{article}
\pdfoutput=1

\usepackage{jheppub}
\usepackage{graphicx}
\usepackage{amsmath}
\usepackage{amssymb}
\usepackage{xspace}
\usepackage{caption}
\usepackage{subcaption}
\usepackage[yyyymmdd,hhmmss]{datetime}
\usepackage[normalem]{ulem}
\usepackage[multiple]{footmisc}
\usepackage{booktabs}
\usepackage{multirow}
\usepackage{pbox}
\usepackage{tabularx}
\usepackage{slashed}
\usepackage{snapshot}

\newcommand{\newc}{\newcommand}
\def\beq{\begin{equation}}
\def\eeq{\end{equation}}
\def\beqn{\begin{eqnarray}}
\def\eeqn{\end{eqnarray}}
\newc{\gluino}{\tilde g}
\newc{\Conep}{\tilde \chi_1^+}
\newc{\Conem}{\tilde \chi_1^-}
\newc{\Cone}{\tilde \chi_1^\pm}
\newc{\Ntwo}{\tilde \chi_2^0}
\newc{\None}{\tilde \chi_1^0}
\newc{\stopp}{\tilde t}
\newc{\stopa}{\tilde t^*}
\newc{\stopone}{\tilde t_1}
\newc{\stopones}{\tilde t_1^*}
\newcommand{\lsim}{\raisebox{-0.13cm}{~\shortstack{$<$ \\[-0.07cm] $\sim$}}~} 
\newcommand{\gsim}{\raisebox{-0.13cm}{~\shortstack{$>$ \\[-0.07cm] $\sim$}}~}

\title{Mono-top Signature from Supersymmetric $t \bar t H$ Channel}

\author[a]{Dorival Gon\c{c}alves,}
\author[a]{Kazuki Sakurai,}
\author[b]{Michihisa Takeuchi}

\affiliation[a]{Institute for Particle Physics Phenomenology, Department of Physics, Durham University, Durham, DH1 3LE, United Kingdom}
\affiliation[b]{Kavli IPMU (WPI), UTIAS, University of Tokyo, Kashiwa, 277-8583, Japan}

\emailAdd{dorival.goncalves@durham.ac.uk}
\emailAdd{kazuki.sakurai@durham.ac.uk}
\emailAdd{michihisa.takeuchi@ipmu.jp}


\abstract{
We point out that a distinctive mono-top signature is present in Natural SUSY scenarios when a scalar top-quark
and higgsinos are almost mass degenerate. This signature originates from a supersymmetric counter part 
of the $t \bar t H$ process, i.e.~$pp \to \tilde t\, t\, \tilde h$. 
Unlike mono-jet signatures exploiting initial
state radiation,  this channel can be regarded as a smoking gun signature of a light stop and higgsinos, 
allowing a direct probe of  the stop and neutralino sectors. 
The production rate of this channel largely depends
on the up-type higgsino components in the neutralinos while the stop sector is sensitive to angular distributions 
of top-quark's decay products.  We develop an optimal search strategy to capture the supersymmetric $t \bar t H$ 
process and find that a high luminosity LHC can probe the stop and higgsino 
sectors with  $m_{\tilde t_1} \lsim 380$ GeV and $m_{\tilde t_1} - m_{\tilde \chi_1^0} \lsim m_W$. 
Additionally, we propose a kinematic variable with which one can measure the stop mixing in this channel.
}

\preprint{
\begin{flushright}
IPPP/16/30,~
IPMU16-0047
\end{flushright}
}

\begin{document}

\maketitle


\section{Introduction}\label{sec:intro}



After the long shut down CERN's Large Hadron Collider (LHC) has resumed colliding protons, 
almost doubling the collision energy to 13 TeV.
With this highest-ever energy, the LHC Run-2 expects to observe the processes with multiple heavy particles 
such as $t \bar t H$ \cite{Aad:2015iha,
Aad:2014lma,
Khachatryan:2014qaa,
Degrande:2012gr,
Ellis:2013yxa,
Nishiwaki:2013cma,
Santos:2015dja,
Li:2015kaa,
Moretti:2015vaa,
Buckley:2015vsa}, $t q H$ \cite{Khachatryan:2015ota,
  Maltoni:2001hu,
  Farina:2012xp,
  Englert:2014pja,
  Kobakhidze:2014gqa,
  Yue:2014tya} and possibly $HH$ \cite{ATL-PHYS-PUB-2015-046,
CMS-PAS-FTR-15-002,
Baur:2002qd,
Dolan:2012rv,
Papaefstathiou:2012qe,
Baglio:2012np,
Baur:2003gp,
Dolan:2012ac,
Goertz:2013kp,
Shao:2013bz,
Gouzevitch:2013qca}.
Observing these processes is not only interesting by its own right 
but also crucial to directly measure the interaction of the Higgs boson with top-quarks and Higgs boson itself.

Another compelling physics target of Run-2 is searches for new physics beyond the Standard Model.
The leading candidate of such models is Supersymmetry (SUSY), in which the gauge hierarchy problem is elegantly solved
due to the underlying symmetry between bosons and fermions.
In the Minimal SUSY Standard Model (MSSM) the bare Higgs mass-squared parameter and the radiative correction to it are given by the mass scales of higgsinos and scalar top-quarks (stops), respectively.
{\it Naturalness}, therefore, requires higgsinos and stops not to be significantly heavier than the gauge boson mass scale, whilst it lefts the rest of the spectrum rather unconstrained.\footnote{
    Except for gluinos.  The gluinos contribute to the radiative corrections to the Higgs mass-squared parameter
    through  renormalisation group evolution  of the stop mass.
    Since the sensitivity of the gluino mass to naturalness is higher order compared to that of  stops and higgsinos,
    in this paper we focus only on light stops and higgsinos. 
}
Indeed, naturalness remains almost intact even if all other SUSY particles are pushed up 
to a few TeV, significantly heavier than their exclusion limit obtained in the Run-1 and early 13 TeV data collected in 2015.
Such a scenario, called {\it Natural SUSY}, has been extensively studied in the literature \cite{Papucci:2011wy,
Hall:2011aa,
Desai:2011th,
Ishiwata:2011ab,
Sakurai:2011pt,
Kim:2009nq,
Wymant:2012zp,
Baer:2012up,
Randall:2012dm,
Cao:2012rz,
Asano:2012sv,
Baer:2012uy,
Evans:2013jna,
Hardy:2013ywa,
Kribs:2013lua,
Bhattacherjee:2013gr,
Rolbiecki:2013fia,
Curtin:2014zua, 
Kim:2014eva,
Papucci:2014rja,
Casas:2014eca,
Katz:2014mba,
Heidenreich:2014jpa,
Mustafayev:2014lqa,
Brummer:2014yua,
Cohen:2015ala,
Hikasa:2015lma,
Barducci:2015ffa,
Drees:2015aeo,
Mitzka:2016lum}.

Reflecting its importance and non-triviality \cite{Plehn:2010st,
Plehn:2011tf,
Han:2012fw,
Kilic:2012kw,
Alves:2012ft,
Bai:2012gs,
Buckley:2013lpa,
Bai:2013ema,
Buckley:2014fqa,
Czakon:2014fka,
An:2015uwa,
Kobakhidze:2015scd,
Macaluso:2015wja,
Cheng:2016mcw}, numerous ATLAS and CMS analyses have been devoted to 
light stop searches.
The exclusion limit on the mass of the lighter stop, $\tilde t_1$, largely depends on its decay modes.
In Natural SUSY lighter neutralinos ($\tilde \chi_{1}^0$ and $\tilde \chi_{2}^0$) and the lighter charginos ($\tilde \chi_1^\pm$) 
are higgsino-like and almost mass degenerate: $m_{\tilde \chi_1^0} \simeq m_{\tilde \chi_2^0} \simeq m_{\tilde \chi_1^\pm}$.
If $\tilde t_1 \to t \tilde \chi_1^0$ is kinematically forbidden ($m_{\tilde t_1} < m_{\tilde \chi_1^0} + m_t$),
the decay mode of $\tilde t_1$ is dominated by
\beq
\tilde t_1 \to b \chi_1^\pm \,.
\eeq
Due to the mass degeneracy between $\tilde \chi_1^\pm$ and $\tilde \chi_1^0$,
the subsequent decay $\tilde \chi_1^\pm \to f \bar f \tilde \chi_1^0$ would not be observable.
ATLAS and CMS have searched for this process in the di-$b$-jet channel 
\cite{Aad:2013ija, Khachatryan:2015wza, ATLAS-CONF-2015-066}. 
Currently, the most stringent bound, $m_{\tilde t_1} \gsim 840$ GeV for $m_{\tilde \chi_1^0} \lsim 200$ GeV, comes from the 13 TeV ATLAS analysis \cite{ATLAS-CONF-2015-066} 
with the integrated luminosity of 3.2 fb$^{-1}$.
However, this limit diminishes if the mass difference $\Delta m_{\tilde t_1 - \tilde \chi_1^0} \equiv m_{\tilde t_1} - m_{\tilde \chi_1^0}$ gets compressed, because the $b$-quarks from the stop decays become soft and undetectable.
For instance, it becomes as  weak as $m_{\tilde t_1} \gsim 300$ GeV if $\Delta m_{\tilde t_1 - \tilde \chi_1^0} \lsim 50$~GeV.

\begin{figure}[!t]
    \begin{center}
          \includegraphics[scale=0.34]{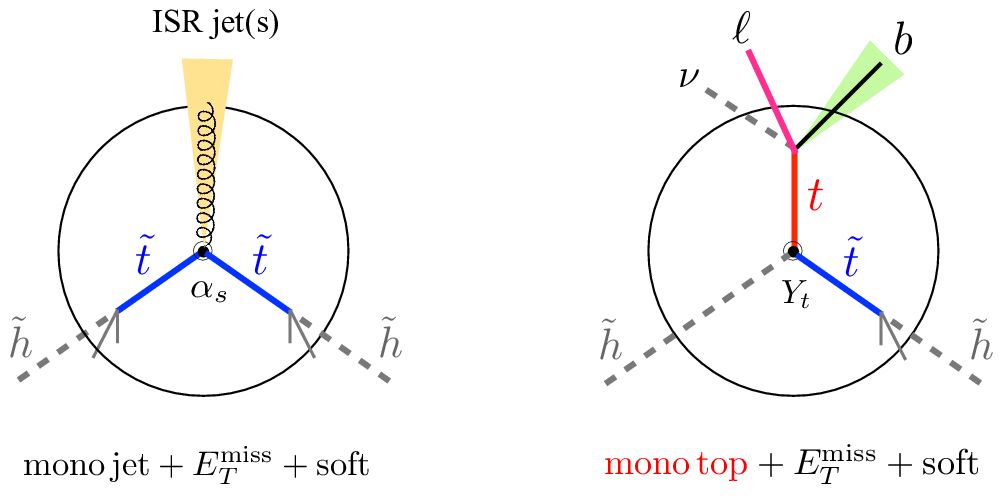}          
    \caption{
    Mono-jet event topologies channel from $\tilde t_1$ pair production (left)
    and mono-top from supersymmetric $t \bar t H$ process (right).  The grey dashed
    lines represent invisible particles, while the thin grey lines represent particles that
    are too soft to be observed. 
    The strong coupling and the top Yukawa coupling are denoted as $\alpha_s$ and $Y_t$, respectively.
    } 
    \label{fig:event_display}
    \end{center}
\end{figure}

The compressed stop-higgsino region can be searched for by exploiting the stop pair production 
associated with hard QCD initial state radiation (ISR).
In such events the system of two stops is boosted recoiling against the high $p_T$ ISR jets, leading to
a mono-jet signature as shown in the left panel of Fig.~\ref{fig:event_display}. 
%
%
%
%
Although the mono-jet channel is useful for discovery, it has some disadvantages.
%

\begin{itemize}
\item 
Since it requires at least one high $p_T$ QCD jet, the cross section is suppressed by 
the QCD coupling, $\alpha_s(\mu)$, approximately at the scale of the $p_T$ cut, $\gsim \mathcal{O}(100)$~GeV.
\item 
There is a large QCD dijet background where one of the jets is badly mismeasured.
Because of this and the above reason, the limit obtained from the mono-jet channel is rather weak:
$m_{\tilde t_1} \gsim 270$ GeV for $\Delta m_{\tilde t_1 - \tilde \chi_1^0} \lsim 15$ GeV 
\cite{Aad:2014nra,Khachatryan:2015wza}.
The limit deteriorates  if the mass difference increases since the $b$-quark from the $\tilde t_1 \to b \tilde \chi_1^\pm$ decay starts to be visible.  
For example, the limit is weakened to $m_{\tilde t_1} \gsim 200$ GeV for $\Delta m_{\tilde t_1 - \tilde \chi_1^0} \gsim 50$ GeV 
\cite{Aad:2014nra, Khachatryan:2015wza}.
\item 
The signal is entirely controlled by QCD interactions, hence the available information is limited. 
For example, even in the presence of an excess, it would be very difficult to find out what types of particles are produced and how they decay as we would only observe the jets from QCD radiation.
\end{itemize}

In this paper we point out that a large collision energy of 13 TeV LHC opens up the possibility of observing 
the stop-top-higgsino production process, $pp \to \tilde t_1 t \tilde \chi^0_{1(2)}$,\footnote{
  We consider both $\tilde t_1^* t \tilde \chi^0_{i}$ and $\tilde t_1 \bar t \tilde \chi^0_{i}$
  but simply write $\tilde t_1 t \tilde \chi^0_{i}$.
}
providing an additional handle for the compressed stop-higgsino region in Natural SUSY.
This process is nothing but a supersymmetric counter part of the $t \bar t H$ process, and analogously to the $t \bar t H$ 
it is crucial to directly probe the interaction between stops and higgsinos.
Because the stop is essentially invisible as its decay products are too soft to be observed in the compressed region, 
the process leads to a distinctive mono-top signature as depicted in the right panel of Fig.~\ref{fig:event_display}.
The mono-top signature has been actively studied mainly in the context of the flavour violating models 
\cite{Aad:2014wza, Khachatryan:2014uma, Davoudiasl:2011fj, Kamenik:2011nb, 
Andrea:2011ws, Alvarez:2013jqa, Agram:2013wda, Boucheneb:2014wza}.
The process discussed in this paper, however, does not belong to this type since the mono-top nature emerges 
due to the kinematics of the stop's decay products.
In contrast to the mono-jet channel, this process has the following advantages.
\begin{itemize}
\item Despite a large mass of the system, the production rate is not too small because 
the stop-top-higgsino interaction is proportional to the top Yukawa coupling, $Y_t$.
\item The QCD multijet background can be controlled by requiring an isolated  lepton 
from top-quark decays. 
\item The process contains rich information on the stop and neutralino sectors.
For example, as will be shown in the next section, the production cross section depends dominantly on
the up-type higgsino components in the neutralinos.\footnote{
    The details of the neutralino sector may also be probed via the $pp \to \tilde q \tilde \chi_1^0$ process
    if squarks are light and $\tilde \chi_1^0$ is gaugino-like \cite{Allanach:2010pp,Binoth:2011xi}.
}
On the other hand, the structure of the stop mixing can be probed by looking at
the kinematic distributions of the $b$-jet and the lepton from the top-quark decay as we will see in section \ref{sec:distribution}.

\end{itemize}

The paper is organised as follows.
In the next section, we study the production cross section of the supersymmetric $t \bar t H$ process
and discuss how the cross section {\it does} and {\it does not} depend on the neutralino and the stop sectors.
In section \ref{sec:analysis}, an optimal search strategy is proposed based on various kinematic distributions of
the signal and background. We derive the 2-$\sigma$ sensitivity
assuming 13 TeV LHC with 3 ab$^{-1}$ of the integrated luminosity.
In section \ref{sec:distribution}, we demonstrate how the stop mixing parameter can be probed 
by looking at the kinematic distributions of the top-quark decay products.
We conclude this paper in section \ref{sec:conclusion}.

\section{Cross Section of the Supersymmetric $t \bar t H$ process}
\label{sec:rates}

\begin{figure}[!t]
    \begin{center}
          \includegraphics[scale=0.38]{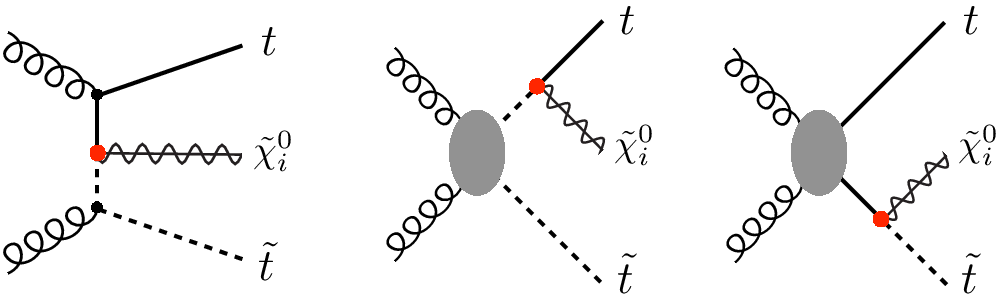}          
    \caption{Representative Feynman diagrams for the supersymmetric $t \bar t H$ process.
    The $q \bar q$ initial states are also possible for the latter two diagrams.
    The red dots denote the stop-top-higgsino interaction.
    The stop propagator in the second diagram has to be far off-shell 
    in our parameter region $m_{\tilde t_1} - m_{\tilde \chi_1^0} < m_W$,
    hence it is clearly separated from the stop pair production. 
    }
    \label{fig:diagrams}
    \end{center}
\end{figure}

Fig.~\ref{fig:diagrams} shows some of the tree-level diagrams contributing to 
the supersymmetric $t \bar t H$ process, i.e.~$pp \to \tilde t_1 t \tilde \chi_i^0$ ($i \in \{1,2\}$).
As mentioned in the previous section, in Natural SUSY scenarios
$\tilde \chi_1^0$ and $\tilde \chi_2^0$ are higgsino-like and almost mass degenerate. Therefore,
both $\tilde t_1 t \tilde \chi_1^0$ and $\tilde t_1 t \tilde \chi_2^0$ processes  contribute to the signal.
In this paper, we focus on the compressed stop-higgsino region, in particular $m_{\tilde t_1} < m_{\tilde \chi_1} + m_W$,
since searches for light stops in this parameter regime are experimentally challenging.
It is worthwhile to note that if the mass difference is larger than $m_t$,
the supersymmetric $t \bar t H$ process cannot easily be distinguished from the
$\tilde t_1$ pair production where one of the stops decays into $t$ and $\tilde \chi_{1(2)}^0$.
The compressed stop-higgsino region studied in this paper does not have such a complication. 
As can be seen in Fig.~\ref{fig:diagrams}, the amplitude is proportional to the stop-top-neutralino vertex
depicted by the red dots, and one can probe the stop and neutralino sectors through this interaction.
%
%
%
%

Before going to the details, 
we define the stop mixing as
\beq
\begin{pmatrix}
\tilde t_1 \\ \tilde t_2 
\end{pmatrix}
= 
\begin{pmatrix}
\cos \theta_{\tilde t} &&  \sin \theta_{\tilde t} \\
- \sin \theta_{\tilde t} &&   \cos \theta_{\tilde t}
\end{pmatrix}
\begin{pmatrix}
\tilde t_R \\ \tilde t_L 
\end{pmatrix}
\eeq
with $m_{\tilde t_1} \le m_{\tilde t_2}$.
The neutralino mass matrix is given by
\beq
M_{\psi} =  
\begin{pmatrix}
M_1 & 0 & - \cos \beta \sin \theta_W m_Z &  \sin \beta \sin \theta_W m_Z \\
0 & M_2 &   \cos \beta \cos \theta_W m_Z & -\sin \beta \cos \theta_W m_Z \\
-\cos \beta \sin \theta_W m_Z &  \cos \beta \cos \theta_W m_Z & 0    & -\mu \\
 \sin \beta \sin \theta_W m_Z & -\sin \beta \cos \theta_W m_Z & -\mu & 0    \\
\end{pmatrix}
\eeq
in the basis of $\psi_a = (\tilde B, \, \tilde W^0, \, \tilde h^0_d, \, \tilde h^0_u)$,
where $\tan\beta$ is the ratio of the vacuum expectation values of the up- and down-type Higgs fields
and $\theta_W$ is the weak mixing angle.
The mass matrix is diagonalised as 
$N M_{\psi} N^{T} = {\rm diag}(m_{\tilde \chi_1^0}, m_{\tilde \chi_2^0}, m_{\tilde \chi_3^0}, m_{\tilde \chi_4^0})$
with $|m_{\tilde \chi_i^0}| \le |m_{\tilde \chi_j^0}|$ for $i < j$,
and $\tilde \chi_i^0 = N_{ia} \psi_a$.
If the electroweak gauginos are decoupled, the lighter two neutralinos become purely higgsino-like (pure higgsino limit) 
and the relevant components of the mixing matrix can be written as
\beq
\begin{pmatrix}
N_{13} & N_{14} \\
N_{23} & N_{24}
\end{pmatrix}
=
\begin{pmatrix}
\frac{1}{\sqrt{2}} & \frac{-1}{\sqrt{2}} \\
\frac{i}{\sqrt{2}} & \frac{i}{\sqrt{2}}
\end{pmatrix} .
\eeq

The stop-top-neutralino interaction is given by
\beq
{\cal L} \supset - \frac{g}{\sqrt{2}} \, \tilde t_1^* \, \sum_i 
\bar{\tilde{\chi}}_i^0 \Big[
    \big( D_{h}^* \sin \theta_{\tilde t} + D_{B} \cos \theta_{\tilde t} \big) P_R
    + \big( D_{h} \cos \theta_{\tilde t} + D_{WB}^* \sin \theta_{\tilde t} \big) P_L
\Big] t 
+ {\rm h.c.} \,
\eeq
with
\beq
D_h
\equiv 
\frac{m_t}{m_W \sin\beta} N_{i4} ,
~~~
D_{B} 
\equiv 
-2 Q_u \tan\theta_W N_{i1},
~~~
D_{WB} 
\equiv 
N_{i2} + (2 Q_u - 1)N_{i1} \tan\theta_W ,
\eeq
where $P_{R(L)}\,= \frac{1 \pm \gamma_5}{2}$ is the chirality projection operator and $Q_u\,= 2/3$ is the electric charge of the top-quark.

In order to parametrise the deviation from the pure higgsino limit,
we define {\it higgsino measure} $\mathcal{R}$ as
\beq
\mathcal{R} \equiv \sigma / \sigma_{\tilde h} \,,
\eeq 
where $\sigma$ is the total cross section of 
the $\tilde t_1 t \tilde \chi_1^0$ and $\tilde t_1 t \tilde \chi_2^0$
processes in the model 
and $\sigma_{\tilde h}$ is that 
for the pure higgsino limit with $\tilde t_1 = \tilde t_R$ and $\sin\beta \simeq 1$.
In the regime where $\tilde \chi^0_{1(2)}$ are higgsino-like ($|N_{i4}| \gg |N_{i1}|,|N_{i2}|$ for $i \in \{1,2\}$) we find approximately
\beq
\mathcal{R} \simeq \frac{|N_{14}|^2 + |N_{24}|^2}{\sin^2\beta} \,.
\label{eq:R}
\eeq
Within this approximation the cross section is independent of the stop mixing
(we will confirm this numerically in section \ref{sec:bounds}) and depends dominantly on
the up-type higgsino components in $\tilde \chi_1^0$ ($N_{14}$) and $\tilde \chi_2^0$ ($N_{24}$)
up to the $1/\sin^2\beta$ factor.\footnote{
  This factor is never significant unless $\tan\beta$ is extremely small.
  For instance $\sin^2\beta = 0.9$, 0.8 and 0.5 for $\tan\beta = 3$, 2 and 1, respectively.
  Moreover, small $\tan\beta$ is not favoured in Natural SUSY scenarios
  since realising $m_h \simeq 125$ GeV becomes even more challenging with light stops.
}  

Eq.~\eqref{eq:R} 
has an important implication.
In the compressed stop-higgsino region $(m_{\tilde t_1} \simeq m_{\tilde \chi_2} \simeq m_{\tilde \chi_1}$)
the mono-top signal rate is determined by $m_{\tilde t_1}$ and $\mathcal{R}$,
whilst the mono-jet signal rate is fixed only by $m_{\tilde t_1}$.
Hence, {\it measuring both mono-jet and mono-top signal rates allows to determine $\mathcal{R}$,
enabling us to directly probe the neutralino sector independently of the details of the stop sector.
}

\begin{figure}[!t]
  \begin{center}
          \includegraphics[scale=0.6]{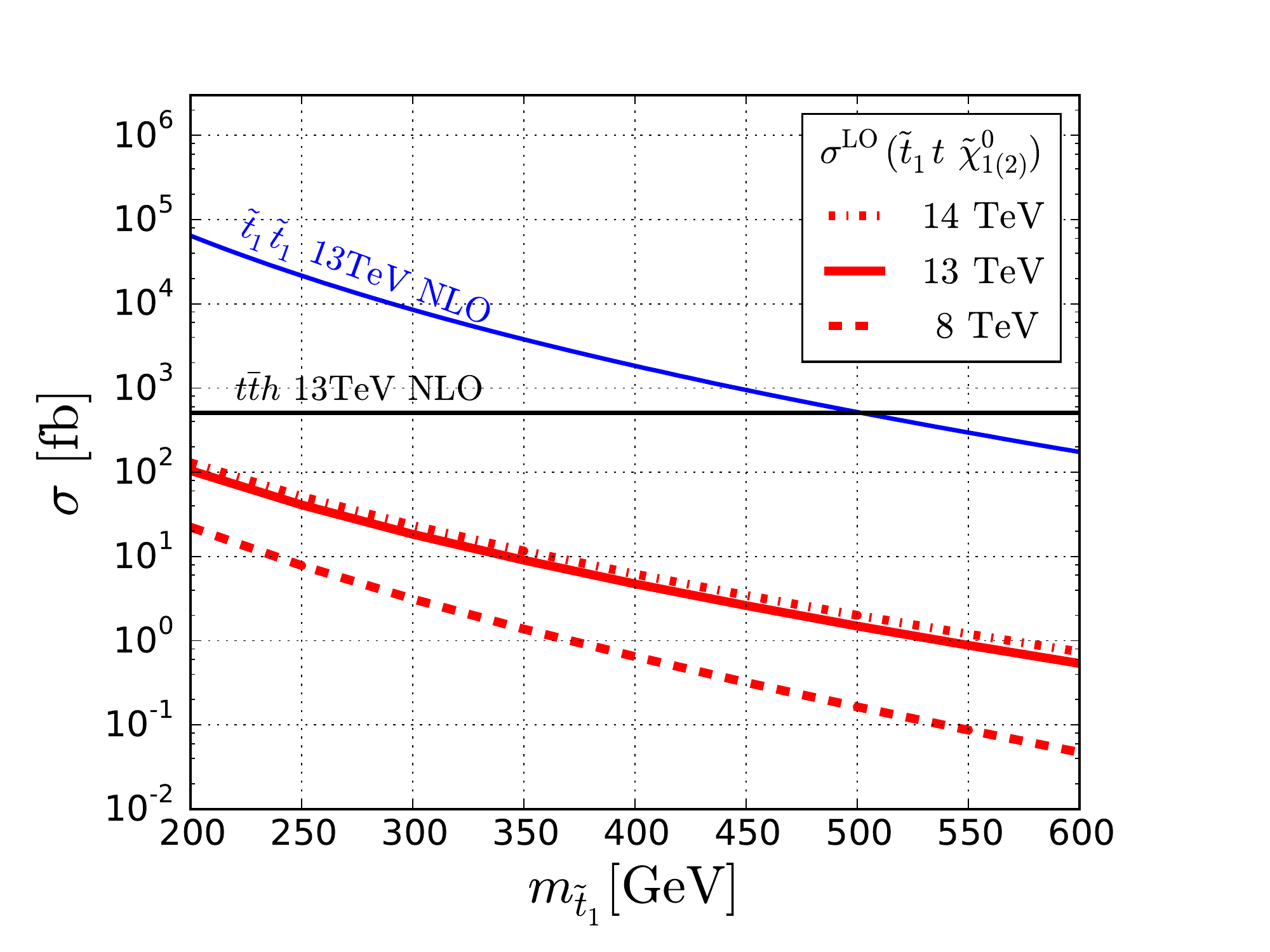}
  \caption{
  LO cross section of the supersymmetric $t \bar t H$ process  
  ($pp \to \tilde t_1 t \tilde \chi^0_{1}$ and $\tilde t_1 t \tilde \chi^0_{2}$ are combined)
   in the pure higgsino limit
  at the 8 (red dashed), 13 (red solid) and 14~TeV (red dashed-dotted) LHC.
  The parameters are fixed as $\Delta m_{\tilde t_1 - \tilde \chi^0_1} = 10$ GeV, $m_{\tilde \chi_2^0} = m_{\tilde \chi^0_1} + 5$ GeV,
  $\cos \theta_{\tilde t} = 1$ and $\tan\beta = 20$.
  These LO cross sections are compared with the NLO cross sections of 
the $\tilde t_1$ pair production (blue solid) and the Standard Model $t \bar t H$ production (black solid)
at the 13 TeV LHC.
  }
   \label{fig:LOxsec}
  \end{center}
\end{figure}

The red curves in Fig.~\ref{fig:LOxsec} show the Leading Order (LO) cross sections of 
the $\tilde t_1 t \tilde \chi^0_{i}$ production ($i=1$ and 2 are combined)
at the 8 (dashed), 13 (solid) and 14~TeV (dashed-dotted)  LHC
in the pure higgsino limit, i.e.~$\mathcal{R} \simeq 1$.
%
%
%
%
We fix 
$\Delta m_{\tilde t_1 - \tilde \chi^0_1} = 10$ GeV, $m_{\tilde \chi_2^0} = m_{\tilde \chi^0_1} + 5$ GeV,
$\cos \theta_{\tilde t} = 1$ and $\tan\beta = 20$ in the calculation.
We use {\tt MadGraph\,5} \cite{mg5} to compute the cross section.
The 13~TeV cross section varies from 105 to 0.53 fb as $m_{\tilde t_1}$ increases from 200 to 600 GeV.
The ratio between the 13 and 8 TeV cross sections ($\sigma_{\rm 13 TeV}/\sigma_{\rm 8 TeV}$) 
is about 5 (10) for $m_{\tilde t_1} = 200$ (600)~GeV. The 14 TeV cross section is not larger than 1.5 times 
the 13 TeV cross section in the range of the plot.

The LO cross section of the supersymmetric $t \bar t H$ process is
compared with the Next-to-Leading Order (NLO) cross sections of 
the $\tilde t_1$ pair production (blue solid) \cite{Beenakker:1997ut, Beenakker:2010nq, Beenakker:2011fu,Goncalves:2014axa, Beenakker:2015rna} 
and the Standard Model $t \bar t H$ production (black solid) \cite{LHCHXSWG}
at the 13 TeV LHC.
The NLO cross section of the $\tilde t_1$ pair production is $\sim 700$ times larger than the LO 
cross section of the $\tilde t_1 t \tilde \chi^0_{i}$ production at $m_{\tilde t_1} = 200$ GeV.
This ratio decreases for larger stop masses 
and becomes $\sim 400$ at $m_{\tilde t_1} = 600$~GeV.
This is because for larger $m_{\tilde t_1}$ (and $m_{\tilde \chi_i^0}$),
the relative importance of the top-quark mass decreases
and the price to produce an extra top-quark diminishes.
The $\tilde t_1 t \tilde \chi^0_{i}$ production at lower stop masses 
has a comparable cross section with that of the Standard Model $t \bar t H$ process.
The former is 105 fb at $m_{\tilde t_1} = 200$~GeV at LO, whereas the latter 508 fb \cite{LHCHXSWG} at NLO.
This is not surprising because these processes share the same coupling due to Supersymmetry.

\section{The Mono-top Search}
\label{sec:analysis}

\subsection{The Search Strategy}

In this section we study various kinematic distributions in the mono-top channel
and develop an optimal search strategy.
We also derive the 2-$\sigma$ sensitivity in the ($m_{\tilde t_1},m_{\tilde \chi^0_1}$) plane
assuming the high luminosity phase 
($\int {\cal L} \,dt = 3$ ab$^{-1}$) of the 13 TeV LHC. 

\begin{figure}[!t]
  \begin{center}
          \includegraphics[scale=0.37]{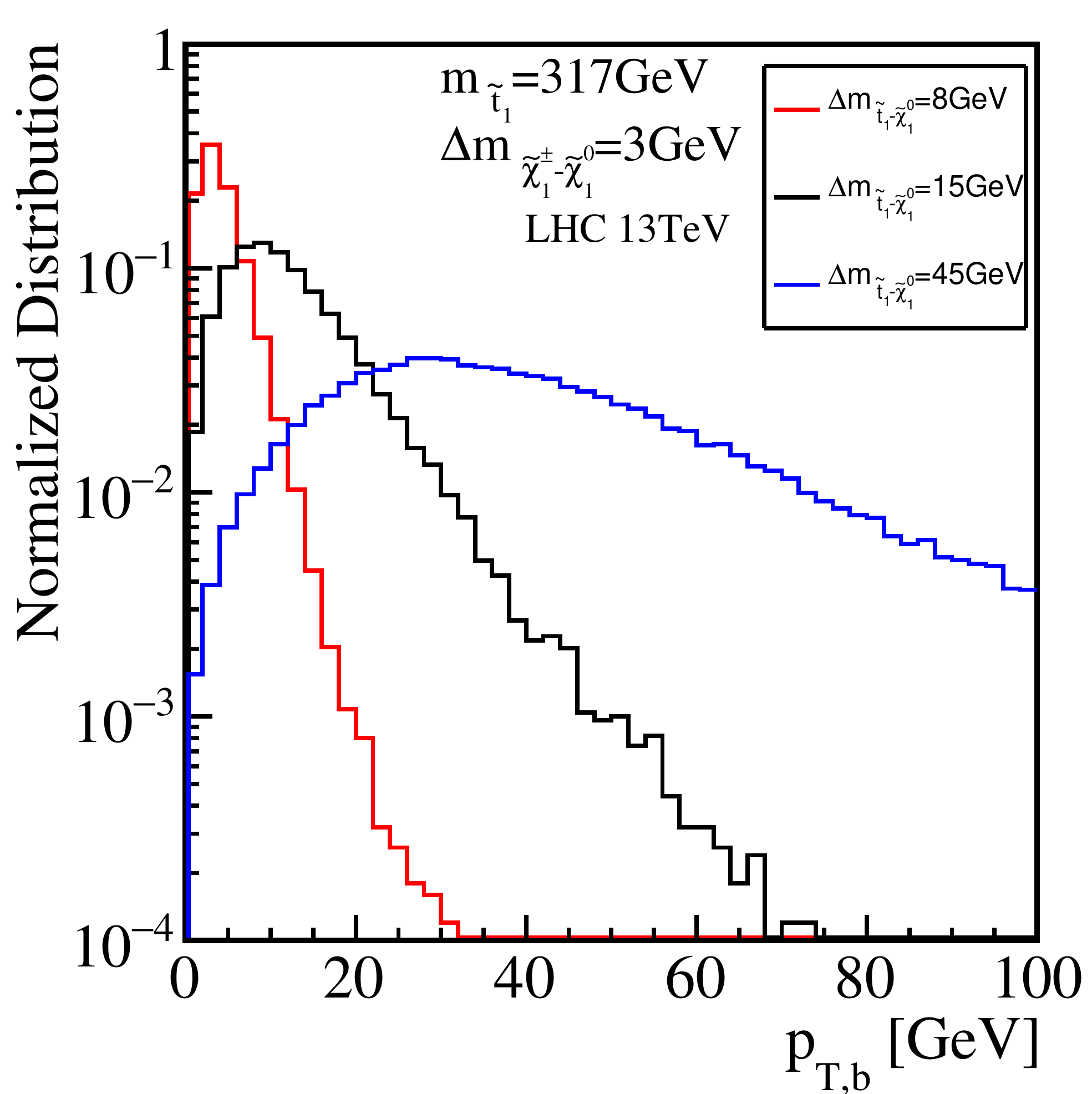}  
          \includegraphics[scale=0.37]{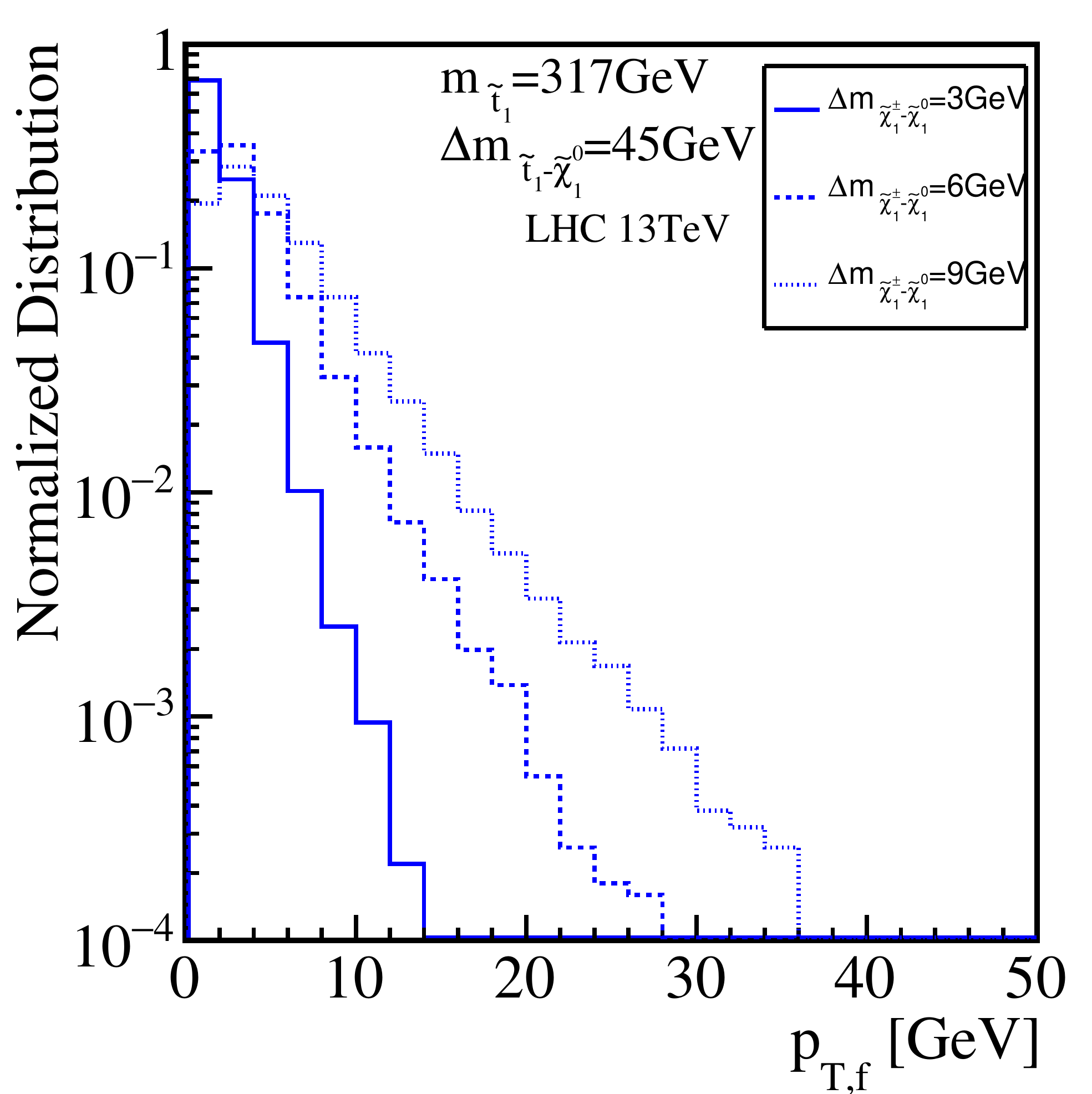}
  \caption{
  Normalised transverse momentum distributions for 
  the $b$-quark from the $\tilde t_1 \to b \tilde \chi_1^\pm$ decay (left panel)
  and the fermions from the subsequent $\tilde \chi_1^\pm \to f \bar f^\prime \tilde \chi_1^0$ decay (right panel)
  at parton level.
  On the left panel we fix the 
  chargino-neutralino mass difference to 
  $\Delta m_{\tilde{\chi}_1^\pm-\tilde{\chi}_1^{0}}=3$~GeV
  and vary the stop-neutralino mass difference as 
  $\Delta m_{\tilde{t}_1-\tilde{\chi}_1^{0}}=8,15$ and 50~GeV, whereas
 on the right panel we fix $\Delta m_{\tilde{t}_1-\tilde{\chi}_1^{0}}=45$~GeV and scan 
 $\Delta m_{\tilde{\chi}_1^\pm-\tilde{\chi}_1^{0}}=3,6$ and 9~GeV.
 We assume $m_{\tilde{t}_1}=317$~GeV for both panels.
   }
  \label{fig:motivation}
  \end{center}
\end{figure}

We begin by looking at the decay products of $\tilde t_1$ in the compressed stop-higgsino region at
parton-level. 
The left panel of Fig.~\ref{fig:motivation} shows the normalised transverse momentum distribution 
for $b$-quarks $p_{Tb}$ from the ${\tilde t_1 \to b \tilde \chi_1^\pm}$ decay. 
We display three distributions with $\Delta m_{\tilde t_1 - \tilde \chi_1^0} = 8$, 15 and 45~GeV
fixing $m_{\tilde t_1} = 317$ GeV and $m_{\tilde \chi_1^\pm} = m_{\tilde \chi_1^0} + 3$ GeV. 
Notice that for a small mass gaps, $\Delta m_{\tilde t_1 - \tilde \chi_1^0} = 8$ and 15 GeV, 
almost all $b$-quarks {\it do not} pass the $p_{Tb} > 30$ GeV cut, whereas for a larger mass difference, 
 ${\Delta m_{\tilde t_1 - \tilde \chi_1^0} = 45}$ GeV, a significant fraction of the $b$-quarks {\it do} pass this selection. 
 
 %
%
%
%

The right panel of Fig.~\ref{fig:motivation} shows the $p_T$ distribution of a fermion (quark or lepton) from the
$\tilde \chi_1^\pm \to f \bar f^\prime \tilde \chi_1^0$ decay.
Differently from the left panel, we now fix $\Delta m_{\tilde{t}_1-\tilde{\chi}_1^{0}}=45$~GeV
and vary the mass difference between $\tilde \chi_1^\pm$ and $\tilde \chi_1^0$ as $\Delta m_{\tilde{\chi}_1^\pm-\tilde{\chi}_1^{0}}=3,6$ and 9~GeV.\footnote{
  In Natural SUSY scenarios the mass difference between $\tilde \chi_1^\pm$ and $\tilde \chi_1^0$
  is smaller than 10 GeV if the electroweak gauginos are heavier than 1 TeV \cite{Bomark:2013nya,Han:2015lma,Badziak:2015qca}.
}
  We observe that $p_{Tf}$ increases on average as $\Delta m_{\tilde{\chi}_1^\pm-\tilde{\chi}_1^{0}}$ increases.
  However, for $\Delta m_{\tilde{\chi}_1^\pm-\tilde{\chi}_1^{0}} \leq 9$ GeV
  the majority of the decay products are always very soft, $p_{Tf} < 10$~GeV.
  We have also checked that the $p_{Tf}$ distribution is almost independent of $\Delta m_{\tilde{t}_1-\tilde{\chi}_1^{0}}$.

These distributions suggest that stop's decay products are soft in the compressed region and unlikely to pass
the standard lepton and jet reconstruction criteria.
In this case, all the visible objects in the final state arise from the top-quark decay (and QCD radiation) as 
illustrated in the right panel of Fig.~\ref{fig:event_display}.
This mono-top feature can be used to efficiently discriminate the signal from backgrounds.

We consider the mono-top signature of the $\tilde t_1 t \tilde \chi_{1(2)}^0$ process with a leptonic top decay, by requiring exactly one isolated lepton ($\ell = e$ and $\mu$) with $p_T > 10$ GeV ($N_\ell(p_T > 10 \,{\rm GeV}) = 1$) and exactly one $b$-tagged jet with $p_T > 30$ GeV ($N_b(p_T > 30 \,{\rm GeV}) = 1$).
To reduce the $t \bar t$ background, we also demand the number of jets with $p_T > 30$ GeV must be less than or equal to three.\footnote{
    Vetoing jets with a $p_T$ much lower than the hard interaction scale may bring a large uncertainty 
    proportional to a logarithm of the ratio of these two scales.
    For a study to understand and reduce this uncertainty, see \cite{Tackmann:2016jyb}. 
}
Our baseline selection cut is thus summarised as
\beq
N_j(p_T > 30\,{\rm GeV}) \le 3,~~~~
N_b(p_T > 30\,{\rm GeV}) = 1,~~~~
N_\ell(p_T > 10\,{\rm GeV}) = 1.
\label{eq:baseline}
\eeq

After these selections, the main backgrounds come from 
$t\bar{t}$ (831\,pb \cite{tt_NNLO}),  
$tW$ (71\,pb \cite{LHCTopWG}),
$tZ$ (0.88\,pb \cite{tz}) and
$W + b \bar b$ (2.55\,pb),
where the numbers in the brackets denote the production rate (before cuts) at 
NNLO+NNLL for $t \bar t$, at LO for $W+b\bar{b}$ and at NLO for all the other processes. 
The $t\bar{t}$ sample is generated with 
{\tt ALPGEN} \cite{Mangano:2002ea} and {\tt Pythia\,6} \cite{pythia} 
and merged up to one jet in MLM matching scheme.
The signal and the other background samples are generated using $\tt MadGraph\,5$ \cite{mg5}
and showered and hadronized with {\tt Pythia\,6}.

The detector effects are included via the {\tt Delphes\,3} package \cite{delphes}.
Jets are defined with the anti-$k_T$ algorithm in {\tt Fastjet} \cite{fastjet1,fastjet2} with $R=0.5$ 
 and required $p_T > 20$ GeV and $|\eta| < 2.5$.
We adopt the $b$-tagging efficiency of $70\,\%$ with $15\,\%$ mistag rate for $c$-quarks and 
$1\,\%$ for light-quarks \cite{btagging,btagging_improv}.
The isolated leptons are defined only within the range of $p_T > 10$ GeV and $|\eta| < 2.4$.

\begin{figure}[!t]
  \begin{center}
          \includegraphics[scale=0.38]{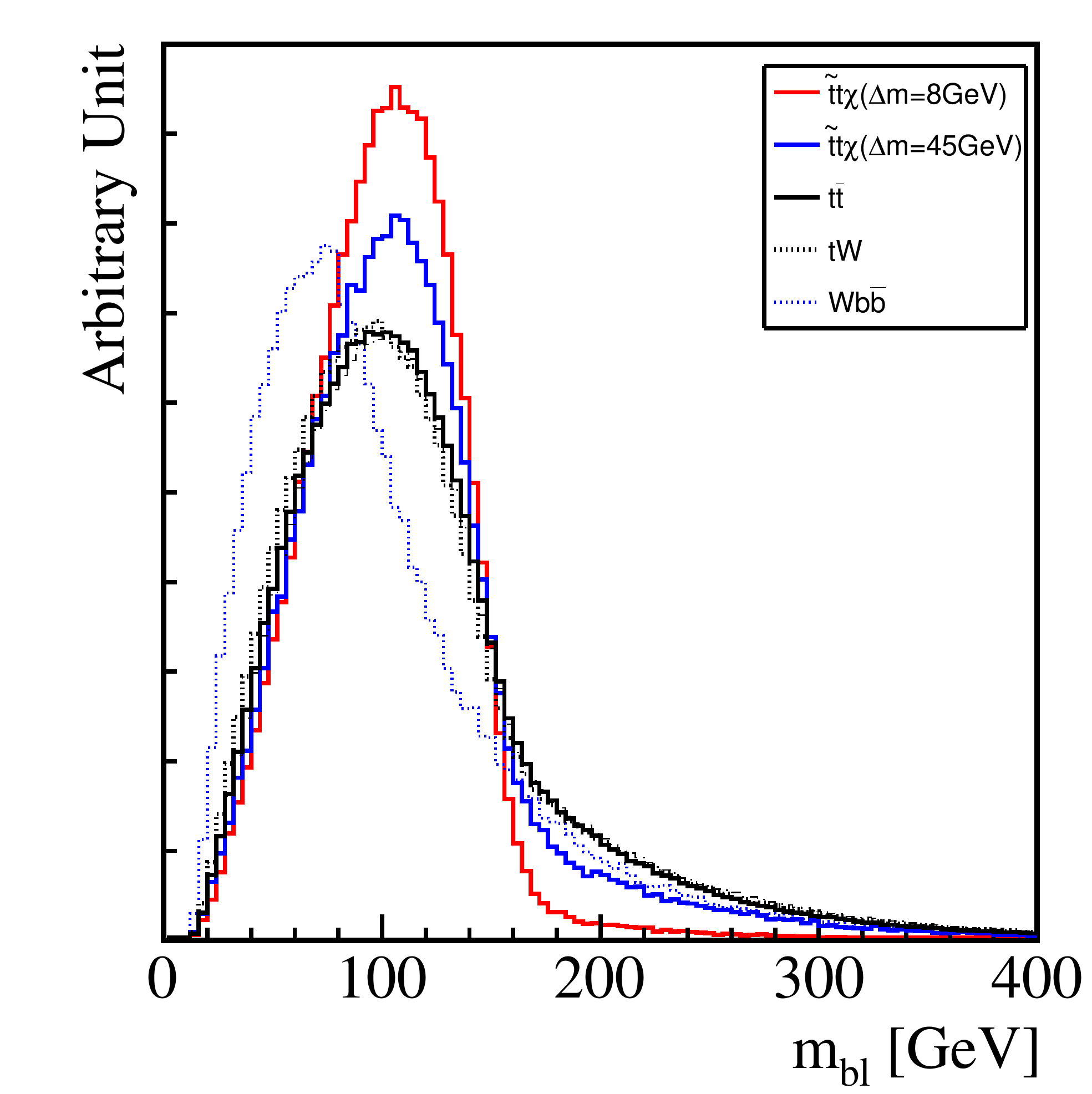}
          \includegraphics[scale=0.38]{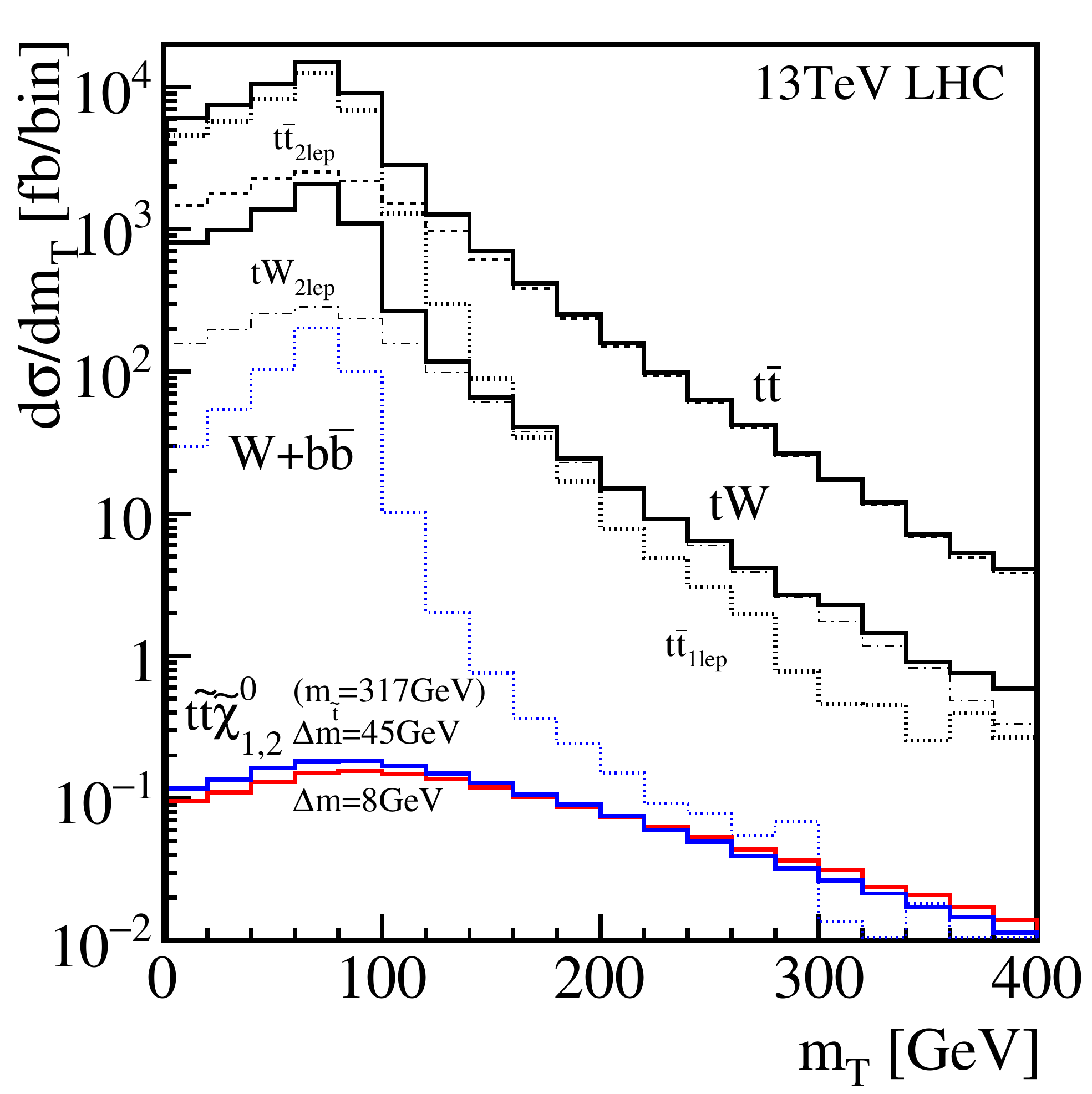}
  \caption{
  Left: Normalised $m_{bl}$ distributions for the signal $\tilde{t}_1 t \tilde{\chi}_{1(2)}^{0}$ 
  with $\Delta m_{\tilde t_1 - \tilde \chi_1^0} = 8$ GeV (red solid) and 45 GeV (blue solid),
  $t\bar{t}$ (black solid), $t W$ (black dotted) and $W+b\bar{b}$ (blue dotted) samples
  after the baseline selection.
  Right: Transverse mass $m_{T}$ distributions after the baseline and $m_{b \ell} < 150$ GeV cuts 
  expected at the 13 TeV LHC.
  The line types and colours are assigned in the same way as in the left panel apart from
  the $t W$, for which the black solid is used.
  In this plot the contributions of the $t \bar t$ and $tW$ where one or two $W$ (and $t$)
  decay(s) leptonically (including $\tau$) are also shown, which are
  $t \bar t_{2l}$ (black dashed), $t \bar t_{1l}$ (black dotted) and $t W_{2l}$ (black dotted-dashed). 
  For both plots the signal points have $m_{\tilde t_1} = 317$ GeV and
  $m_{\tilde \chi_1^\pm} = m_{\tilde \chi_2^0} = m_{\tilde \chi_1^0} + 3$ GeV.
  }
  \label{fig:mT}
  \end{center}
\end{figure}

Using the detector-level samples after applying the baseline selection Eq.~\eqref{eq:baseline},
we now show the distributions of the invariant mass of the $b$ and $\ell$ ($\ell = e, \mu$)
in the left panel of Fig.~\ref{fig:mT}.
%
%
%
%
As can be seen, the signal presents a Jacobian peak structure at $m_{b \ell} \sim 130$ GeV
and most of the signal events fall below $150$ GeV.
This structure is expected if the $b$ and $\ell$ are originated from the same top-quark decay.
Unlike the signal, the $m_{b \ell}$ distributions for $t \bar t$ and $W + b \bar b$
exhibit large tails exceeding 150 GeV.
For $t \bar t$, this tail typically comes from the events where the $b$ and $\ell$
come from different top-quark decays. For $W + b \bar b$, the Jacobian peak structure is not 
expected at the first place, since there is no top-quark in the event. To exploit this feature we impose 
\beq
m_{b \ell} < 150~{\rm GeV} \,.
\label{eq:mblcut}
\eeq

Another variable that is useful to control the background is 
the transverse mass of the lepton and the missing energy vector:
$m_T= \sqrt{ 2 p_{T\ell} E_T^{\rm miss} (1 - \cos\phi_{\ell, {E}_T^{\rm miss}}) }$.
If the lepton and the missing energy are originated from a single $W$ boson, 
this variable is kinematically bounded from above by $m_W$.
This is the case for $W+b \bar b$ and the fraction of the $t \bar t$ events where one of the tops decays
 hadronically and the other leptonically  $t \bar t_{1l}$ (including $\tau$).
The right panel of Fig.~\ref{fig:mT} shows the $m_T$ distribution for the 13 TeV LHC with 3 ab$^{-1}$ of data.
As expected, the $m_T$ distributions in the $W + b \bar b$ and $t \bar t_{1l}$ samples 
sharply drop above $m_T \sim m_W$.
We require  
\beq
m_T > 100 ~ {\rm GeV} \,
\label{eq:mTcut}
\eeq
to further suppress these backgrounds.
Above this threshold the dominant backgrounds become $t \bar t$ and $tW$ where all $W$s and tops decay leptonically
 (including $\tau$), respectively denoted by $t \bar t_{2l}$ and $t W_{2l}$.

\begin{figure}[!t]
  \begin{center}
          \includegraphics[scale=0.4]{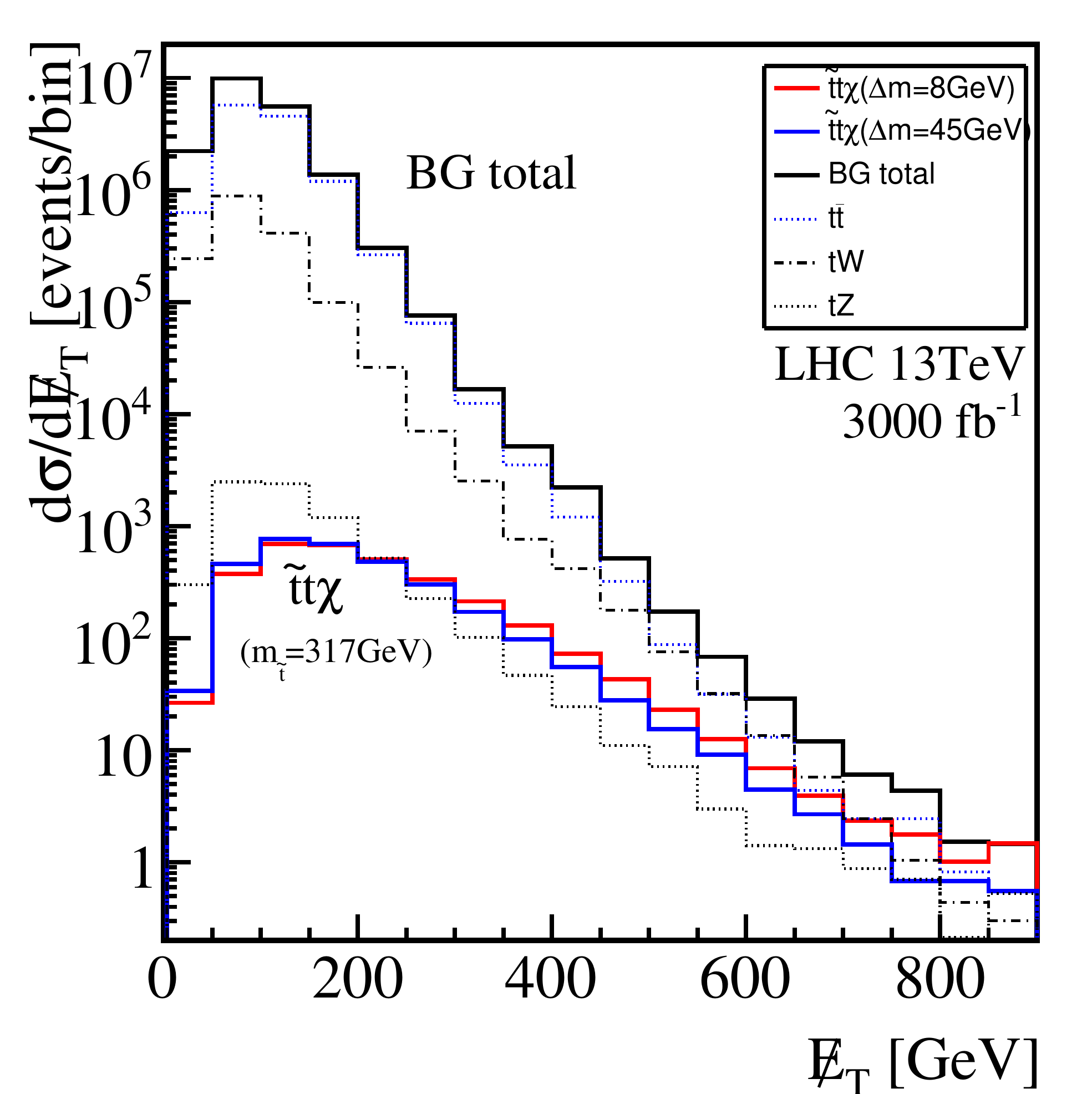}
  \caption{
  Missing energy distribution $\slashed{E}_T$ 
  for the signal $\tilde{t}_1 t \tilde{\chi}_{1(2)}^{0}$ 
  with $\Delta m_{\tilde t_1 - \tilde \chi_1^0} = 8$ GeV (red solid) and 45 GeV (blue solid)
  and the total background (black solid)
  after the selection cuts Eqs.~\eqref{eq:baseline}, \eqref{eq:mblcut} and \eqref{eq:mTcut}.
  The breakdowns of the total background are also shown: 
  $t\bar{t}$ (blue dotted), $t W$ (black dashed-dotted) and $tZ$ (black dotted).
  The distributions are normalised to the expected events expected at the 13 TeV LHC with 3 ab$^{-1}$ integrated luminosity.
  The signal points are assumed to have $m_{\tilde t_1} = 317$ GeV and
  $m_{\tilde \chi_1^\pm} = m_{\tilde \chi_2^0} = m_{\tilde \chi_1^0} + 3$ GeV.
  The last bin is an overflow bin.  
  }
  \label{fig:met}
  \end{center}
\end{figure}

In Fig.~\ref{fig:met}, we display the missing energy distribution for the signal and background samples after
imposing the above selection cuts Eqs.~\eqref{eq:baseline}, \eqref{eq:mblcut} and \eqref{eq:mTcut}.  
%
%
%
%
The $E_T^{\rm miss}$ distribution falls faster for the  total background  than for the signal. We exploit this fact by defining  three
 signal regions (SR1, SR2 and SR3) that correspond to different  missing energy selections 
\beq
E_T^{\rm miss}/{\rm GeV} > 450 ~({\rm SR1}),~500~({\rm SR2}),~550~({\rm SR3}) \,.
\eeq

A detailed cut-flow table showing the number of signal and background events
at a high luminosity LHC with $\sqrt{s}=13$ TeV and $\int \mathcal{L} \, dt =3$ ab$^{-1}$
is presented in Table~\ref{tab:cutflow}. 
Three benchmark points are examined for signal:
$(m_{\tilde t_1}, m_{\tilde \chi_1^0})/{\rm GeV} = (317,\,309)$, (317,\,272) and (342,\,334),
where the remaining parameters are fixed to 
$m_{\tilde \chi_1^\pm} = m_{\tilde \chi_2^0} = m_{\tilde \chi_1^0} + 3$ GeV,
$\tan \beta = 20$ and $\cos \theta_{\tilde t} = 1$.
%
%
\begin{table}[t!]
\centering
\resizebox{\textwidth}{!}{
\begin{tabular}{c||r||r|r|r||r|r|r}
\hline             
  Process    &  $\sigma$  &       Baseline  &  $m_{b\ell} < 150$  &  $m_T > 100$  &  SR1  & SR2  &  SR3 \\
\hline \hline
$t \bar t$   &   831\,pb  &  $206 \cdot 10^6$  &  $165. \cdot 10^6$   &  $17.7 \cdot 10^6$  &  463.3  &  142.6  &  55.2 \\ 
$t W$        &    71\,pb  &  $26.2 \cdot 10^6$  &  $20.7 \cdot 10^6$  &  $1.68 \cdot 10^6$  &  308.5  &  130.9  &  55.5   \\
$t Z$        &  0.88\,pb  &  $22.8 \cdot 10^3$  &  $21.6 \cdot 10^3$  &  $7.3 \cdot 10^3$   &  26.1  &  15.1  &  8.0   \\ 
$W+b \bar b$ &    2.55\,pb  &  $1.82 \cdot 10^6$     &  $1.51 \cdot 10^6$     &  $42.3 \cdot 10^3$     &  5.9   &  2.8  &  1.4  \\ 
\hline
BG total     &   903\,pb  &  $226 \cdot 10^6$   &  $41.1 \cdot 10^6$  &  $19.4 \cdot 10^6$  &  803.8   &  291.4  &  120.1  \\ 
\hline \hline
\multirow{2}{*}{BP(317,\,309)} & \multirow{2}{*}{23.7\,fb} & \multirow{2}{*}{5883} & \multirow{2}{*}{5491} & \multirow{2}{*}{3387} &  109 & 61.4 & 35.0 \\
             &            &                     &                     &                     & (3.8,\,0.13) &  (3.6,\,0.21) & (3.2,\,0.29)  \\
\hline
\multirow{2}{*}{BP(317,\,272)} & \multirow{2}{*}{30.8\,fb} & \multirow{2}{*}{6522} & \multirow{2}{*}{5491} & \multirow{2}{*}{3123} &  60.2 & 34.9 & 19.1 \\
             &            &                     &                     &                     & (2.1,\,0.07) &  (2.0,\,0.12) & (1.7\,0.16)  \\             
\hline             
\multirow{2}{*}{BP(342,\,334)} & \multirow{2}{*}{16.7\,fb} & \multirow{2}{*}{4119} & \multirow{2}{*}{3834} & \multirow{2}{*}{2395} &  84.0 & 46.8 & 26.8 \\
             &            &                     &                     &                     & (3.0,\,0.10) &  (2.7,\,0.16) & (2.4,\,0.22)  \\
\hline             
\end{tabular}}
\caption{Number of signal and background events assuming a high luminosity LHC with $\sqrt{s}=13$ TeV and $\int \mathcal{L} \, dt =3$ ab$^{-1}$. 
We present results for three signal benchmark points:
$(m_{\tilde t_1}, m_{\tilde \chi_1^0})/{\rm GeV} = (317,\,309)$, (317,\,272) and (342,\,334).
The remaining parameters are fixed to $m_{\tilde \chi_1^\pm} = m_{\tilde \chi_2^0} = m_{\tilde \chi_1^0} + 3$ GeV,
$\tan \beta = 20$ and $\cos \theta_{\tilde t} = 1$.
We assume that the higher order corrections to the signal tantamount to a factor $K_ {NLO}=1.5$.
\vspace{0.5cm}
}
\label{tab:cutflow}
\end{table}
%
%
%
The two numbers in the brackets displayed for signal  in the last three columns are $S/\sqrt{B}$ and $S/B$, respectively.
We assume that the higher order corrections to the signal tantamount to a factor $K_ {NLO}=1.5$.\footnote{We notice that the literature 
does not  provide  higher order corrections to the considered signal process. As we consider all the main backgrounds at least at NLO and given the 
similarities between the signal and stop pair production, we assume a similar NLO K-factor. We indicate however the importance of the 
precise NLO rate determination for future studies.}
Notice that we can achieve with this analysis $S/\sqrt{B} \sim 2 - 3$ with $S/B \sim 0.1-0.2$ for $m_{\tilde t_1} \sim 310 - 340$~GeV.

\subsection{The Expected Performance}
\label{sec:bounds}

We now compare the  signal and background in the signal region and 
derive the 2-$\sigma$ sensitivity at a high luminosity LHC with $\sqrt{s} = 13$ TeV and $\int {\cal L}\,dt = 3$ ab$^{-1}$.
We present the sensitivity in a 2D parameter plane ($m_{\tilde t_1}, m_{\tilde \chi_1^0}$)
assuming $m_{\tilde \chi_1^\pm} = m_{\tilde \chi_2^0} = m_{\tilde \chi_1^0} + 3$~GeV
and do not consider the contribution from $\tilde t_2$.
We also consider two extreme cases: $\tilde t_1 = \tilde t_L$ and $\tilde t_1 = \tilde t_R$.

\begin{figure}[!t]
    \begin{center}
          \includegraphics[scale=0.5]{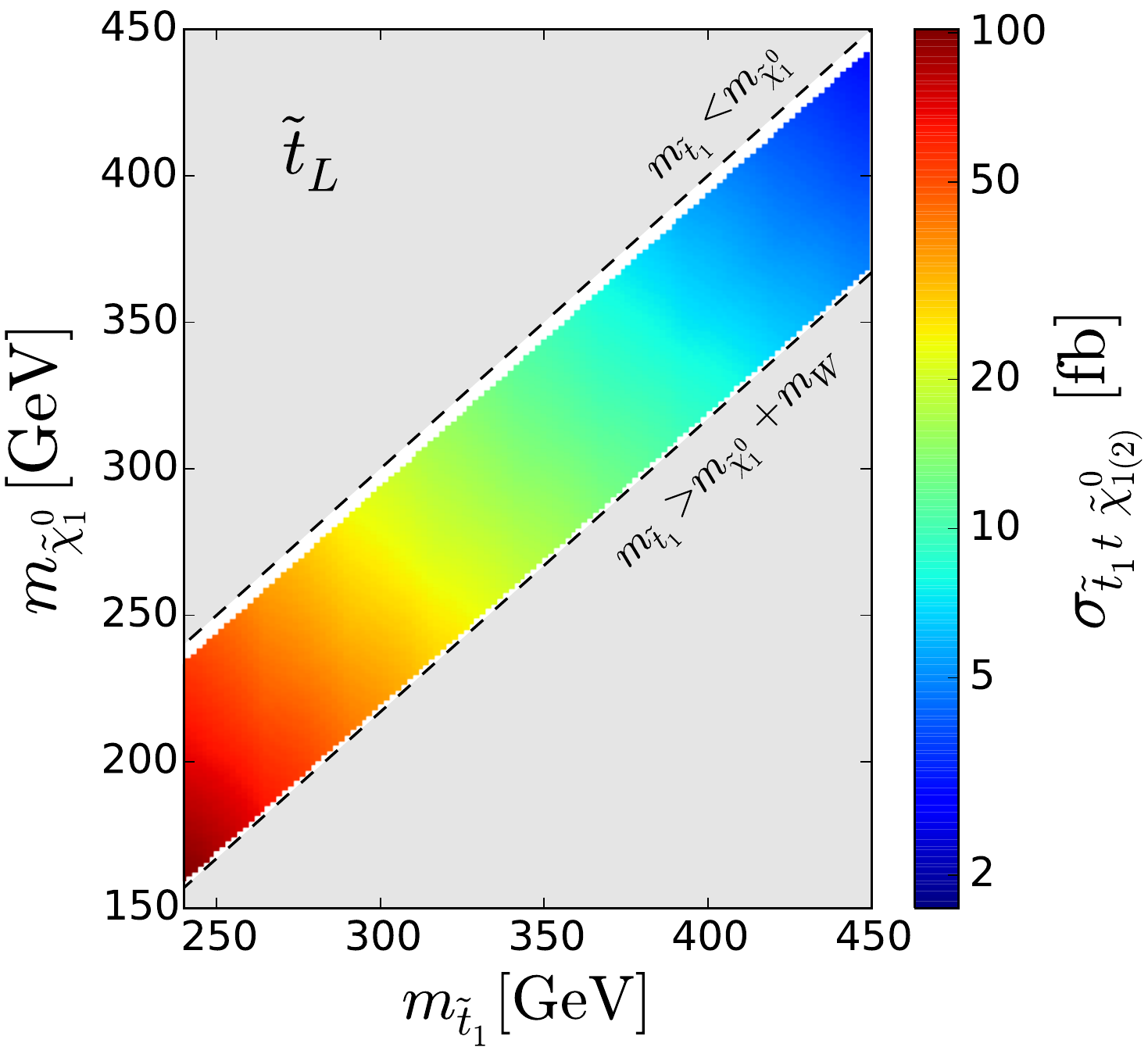}
          \includegraphics[scale=0.5]{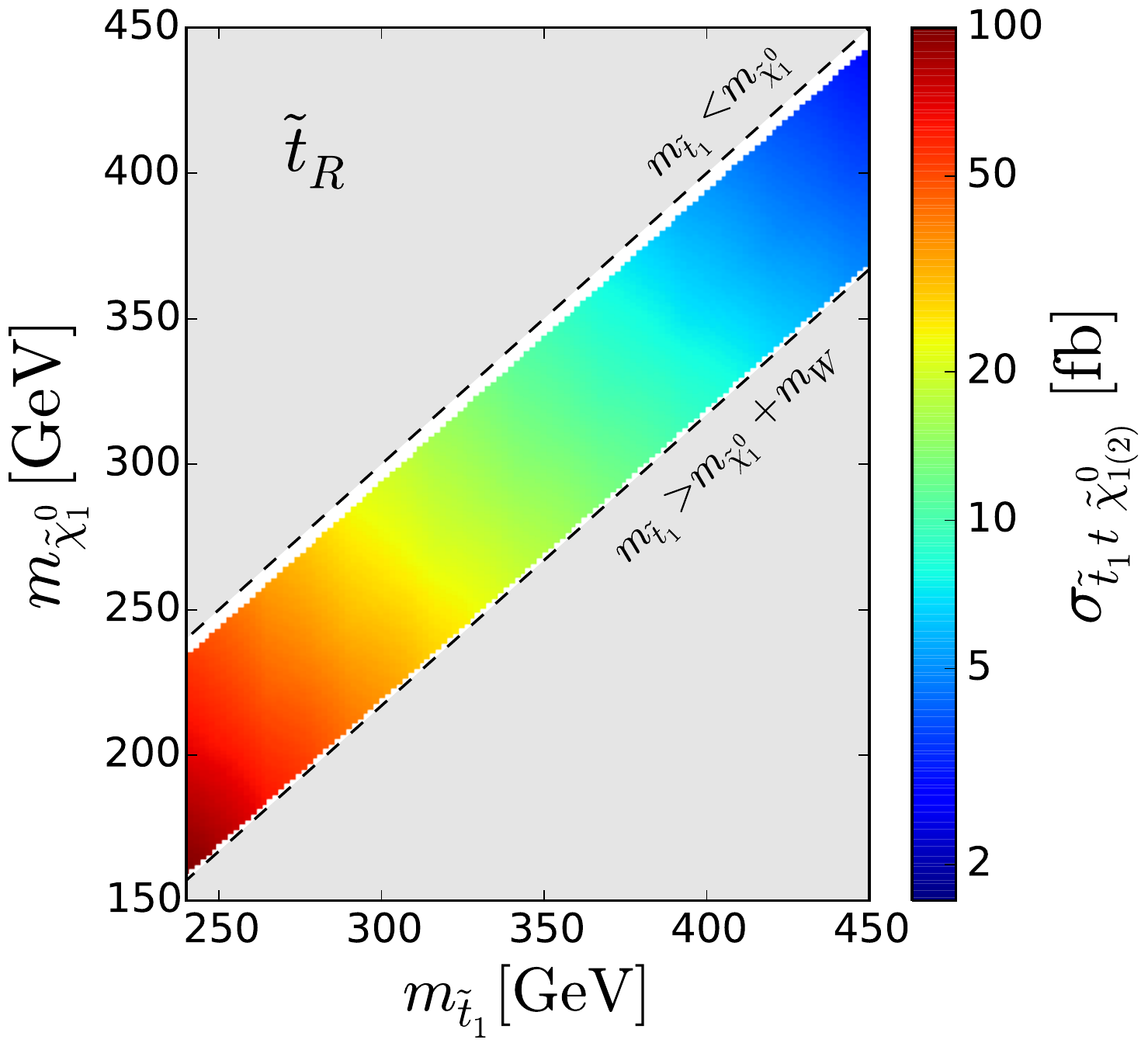}
    \caption{
    The LO cross section in the $(m_{\tilde t_1}, m_{\tilde \chi_1^0})$ plane
    for the $\tilde t_1 = \tilde t_L$ (left) and $\tilde t_1 = \tilde t_R$ (right) cases.
    }
    \label{fig:xsec2d}
    \end{center}
\end{figure}

We first display the LO cross section of the signal in the ($m_{\tilde t_1}, m_{\tilde \chi_1^0}$) plane
in Fig.~\ref{fig:xsec2d} for the $\tilde t_L$ (left panel) and $\tilde t_R$ (right panel) cases.
In the calculation we take the pure higgsino limit for the neutralino mixing 
and $\tan\beta = 20$, i.e.~$\mathcal{R} \simeq 1$.
%
%
%
%
One can see that the cross section 
decreases as either $m_{\tilde t_1}$ and $m_{\tilde \chi_1^0}$ increases.
This is contrasted with the $\tilde t_1$ pair production,
where the cross section depends only on $m_{\tilde t_1}$.
As suggested in Eq.~\eqref{eq:R}, the cross section is almost unchanged 
between the $\tilde t_L$ and $\tilde t_R$ cases.

\begin{figure}[!t]
    \begin{center}
          \includegraphics[scale=0.5]{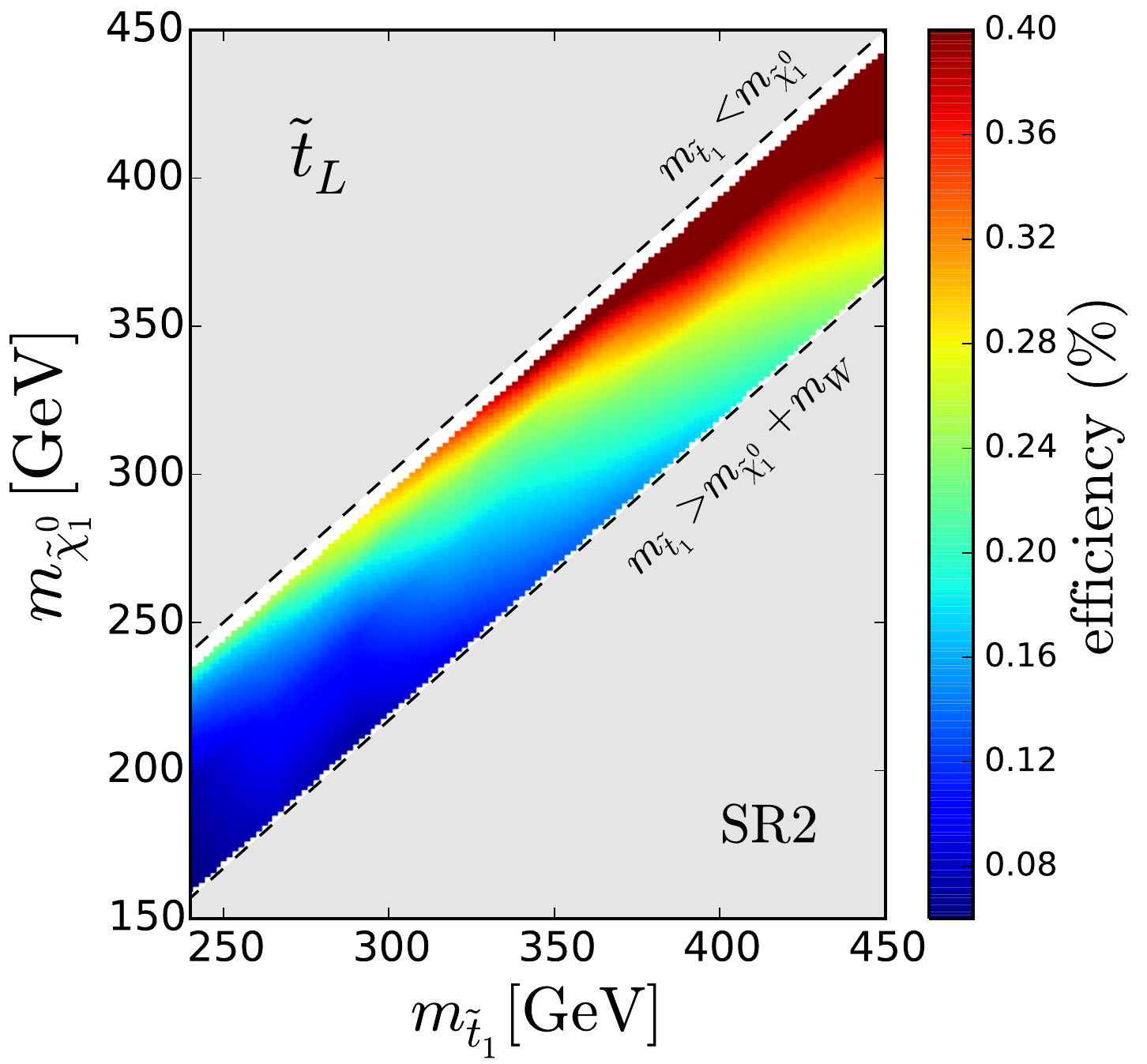}
          \includegraphics[scale=0.5]{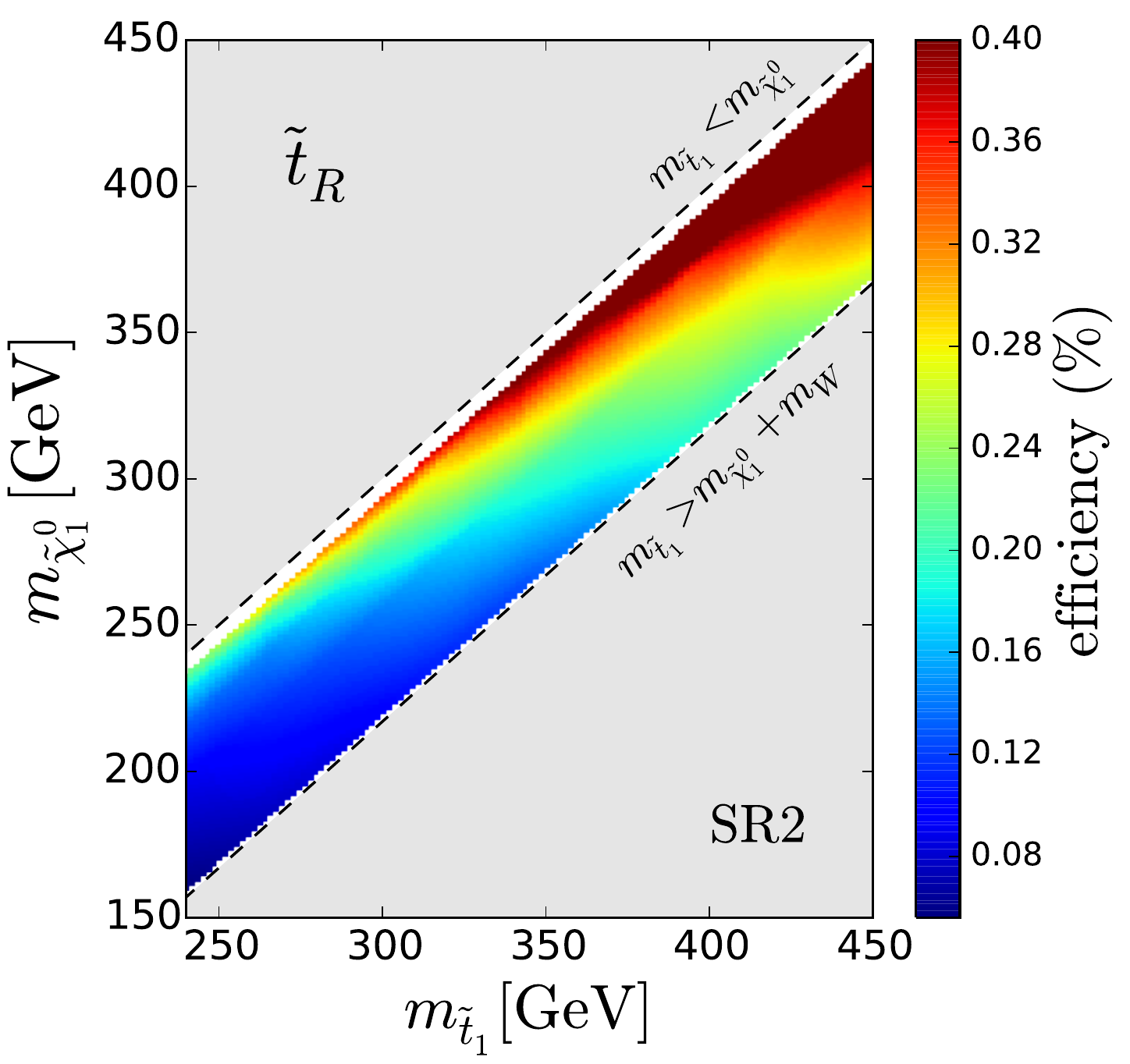}
    \caption{
    The signal efficiency of SR2 in the $(m_{\tilde t_1}, m_{\tilde \chi_1^0})$ plane
    for the $\tilde t_1 = \tilde t_L$ (left) and $\tilde t_1 = \tilde t_R$ (right) cases.    
    }
    \label{fig:eff}
    \end{center}
\end{figure}

We now look how the signal efficiency changes across the 
($m_{\tilde t_1}, m_{\tilde \chi_1^0}$) plane.
As an example, we display the signal efficiency of SR2
in Fig.~\ref{fig:eff} for the $\tilde t_L$ (left panel) and $\tilde t_R$ (right panel) cases.
%
%
%
%
As can be seen, the efficiency varies from 0.08\,\% to $\gsim 0.4 \,\%$
in the region of the plots.
The efficiency is smaller for larger mass difference, $\Delta m_{\tilde t_1 - \tilde \chi_1^0}$.
For larger $\Delta m_{\tilde t_1 - \tilde \chi_1^0}$ 
the $b$-quark from the $\tilde t_1$ decay becomes harder and more visible,
with which the event more likely fails to pass the $N_b = 1$ and $N_j \leq 3$ cuts.
We also observe that the efficiency increases for larger $m_{\tilde t_1}$. 
Since the interaction scale is proportional to the mass of the system,
the typical momentum scale of $\tilde t_1$, $t$ and $\tilde \chi_1^0$ becomes larger as $m_{\tilde t_1}$ increases.
With those high $p_T$ objects, events are more likely to pass the $m_T$ and the missing energy cuts. 
The efficiencies are almost the same for the $\tilde t_L$ and $\tilde t_R$ cases.
This suggests that our search strategy works independently of the details of the stop mixing.

\begin{figure}[!tbh]
    \begin{center}
          \includegraphics[scale=0.55]{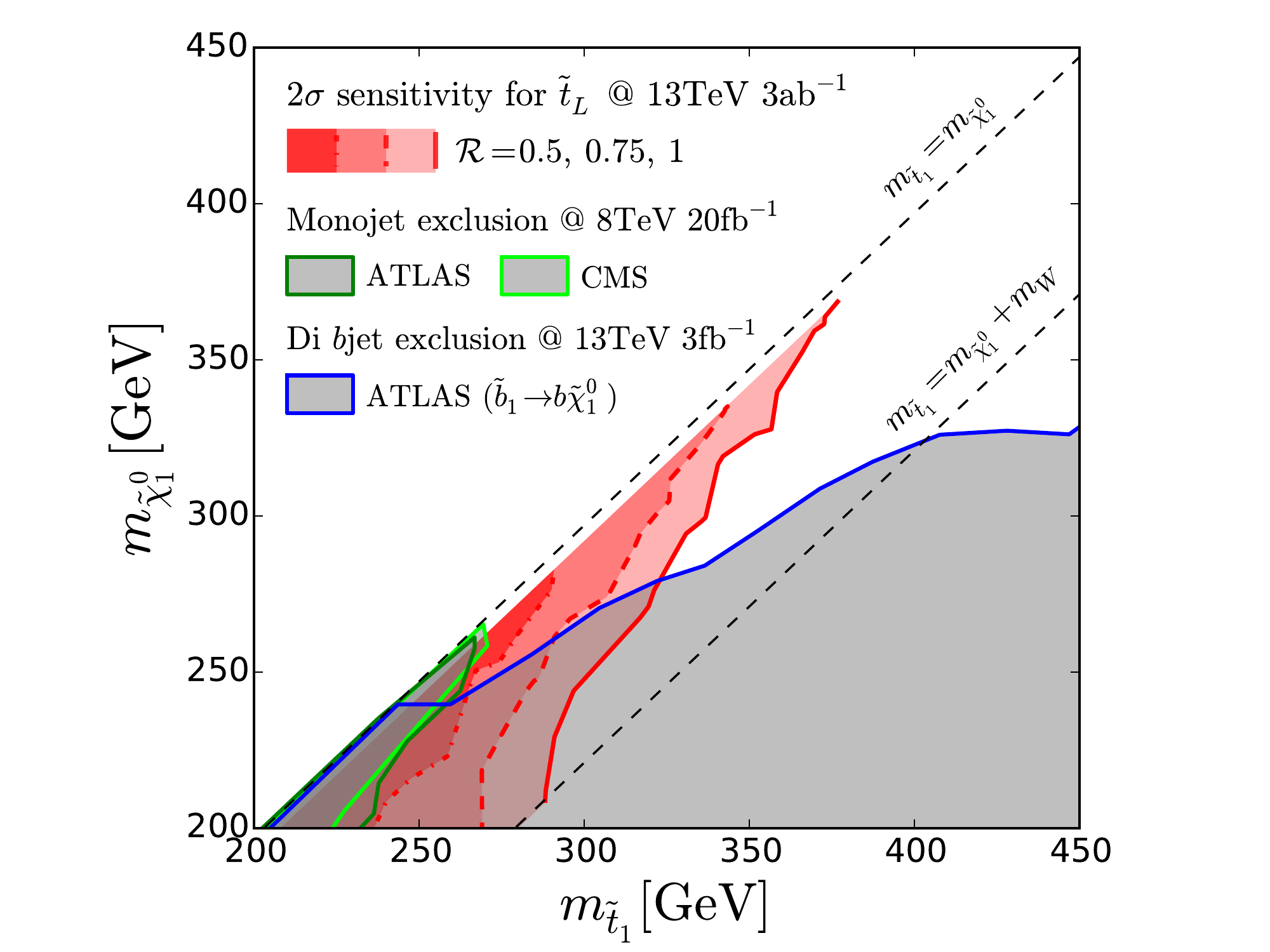}
          \vspace{0.3cm}          
          \includegraphics[scale=0.55]{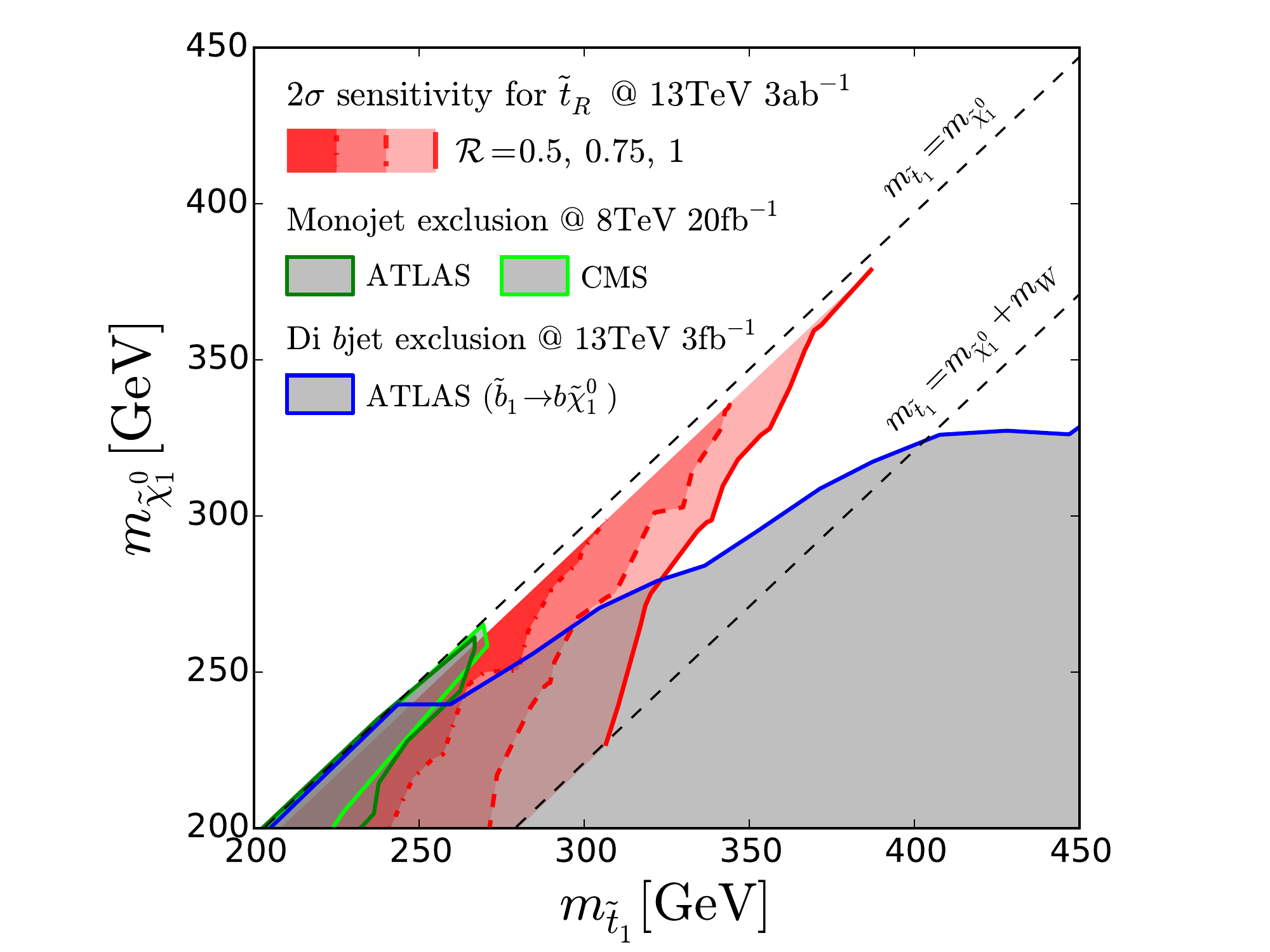}          
    \caption{
The 2-$\sigma$ sensitivities expected at the 13 TeV high luminosity LHC with $\int {\cal L}\,dt = 3$ ab$^{-1}$ 
for $\mathcal{R}= 0.5$ (dark-pink), 0.75 (medium-pink) and 1 (light-pink)
for the $\tilde t_L$ (top) and $\tilde t_R$ (bottom) cases.
In deriving these sensitivities, only the signal regions
with more than three expected signal events and $S/B > 0.1$ are considered
at each parameter point and $\mathcal{R}$.
The signal region with the largest $S/\sqrt{B}$ is then used to derive the sensitivity.  
The current 95\,\% CL excluded region is filled by grey.
The region surrounded by the blue curve is obtained from the 13 TeV ATLAS di-$b$-jet analysis with $\int {\cal L}\,dt = 3.2$ fb$^{-1}$
\cite{ATLAS-CONF-2015-066}.
The regions with dark and light green boundaries are excluded by the ATLAS \cite{Aad:2014nra} and CMS \cite{Khachatryan:2015wza} 
mono-jet searches with Run-1 data corresponding to $\int {\cal L}\,dt \simeq 20$ fb$^{-1}$.
}
    \label{fig:sensitivity}
    \end{center}
\end{figure}

We now show the 2-$\sigma$ sensitivity expected at the 13 TeV LHC with $\int {\cal L}\,dt = 3$ ab$^{-1}$
by the dark-, medium- and light-pink regions in Fig.~\ref{fig:sensitivity}, corresponding to $\mathcal{R} = 0.5$, 0.75 and 1,  respectively.
%
%
%
%
The top and bottom panels are for the $\tilde t_L$ and $\tilde t_R$ cases.
In deriving these sensitivities, only the signal regions
with more than three expected signal events and $S/B > 0.1$ are used
at each parameter point and $\mathcal{R}$.
We then select the signal region that has the largest $S/\sqrt{B}$.  
The most sensitive signal region for each parameter point and $\mathcal{R}$ is given 
in Appendix \ref{sec:srbest}.

We also overlay the current 95\,\% CL exclusion limit for the $\tilde t_1 \to b \chi_1^0$ topology 
with $m_{\tilde \chi_1^\pm} = m_{\tilde \chi_0^\pm}$ by grey regions.
The region surrounded by the blue curve is the 95\,\% CL excluded region by the ATLAS di-$b$-jet search 
\cite{ATLAS-CONF-2015-066} using early 13 TeV data with 3.2 fb$^{-1}$.
ATLAS interprets their analysis in the $\tilde b_1$ production with $\tilde b_1 \to b \tilde \chi_1^0$
and derived the excluded region in the ($m_{\tilde b_1}, m_{\tilde \chi_1^0}$) plane.
Since the production cross section and the decay kinematics are the same between
this $\tilde b_1$ model and the $\tilde t_1$ pair production with $\tilde t_1 \to b \tilde \chi_1^\pm$ at 
$m_{\tilde \chi_1^\pm} = m_{\tilde \chi_1^0}$ and $m_{\tilde t_1} = m_{\tilde b_1}$,
we simply use the $\tilde b_1$ excluded region for $\tilde t_1$ by identifying $m_{\tilde t_1} = m_{\tilde b_1}$.
In realistic models with higgsino-like $\tilde \chi_1^0$, $m_{\tilde \chi_1^\pm}$ is a few GeV larger than
$m_{\tilde \chi_1^0}$.
We therefore believe that the presented exclusion region in Fig.~\ref{fig:sensitivity} 
is slightly aggressive in the compressed stop-higgsino region
since the $b$-quark from $\tilde t_1 \to b \tilde \chi_1^\pm$ is softer 
compared to that from $\tilde b_1 \to b \tilde \chi_1^0$ at the 
same $m_{\tilde \chi_1^0}$ and $m_{\tilde t_1} = m_{\tilde b_1}$.
The other two regions with dark and light green boundaries are the 95\,\% CL excluded region
by mono-jet searches by ATLAS \cite{Aad:2014nra} and CMS \cite{Khachatryan:2015wza}
based on Run-1 data.

One can see from Fig.~\ref{fig:sensitivity} 
that the mono-top search is sensitive for smaller $\Delta m_{\tilde t_1 - \tilde \chi_1^0}$.
This is expected since the $b$-jet from $\tilde t_1 \to b \tilde \chi_1^\pm$ becomes
visible for larger $\Delta m_{\tilde t_1 - \tilde \chi_1^0}$, making the event difficult to pass
the $N_b = 1$ and $N_j \leq 3$ cuts.
The reach of the 2-$\sigma$ sensitivity largely depends on the higgsino measure, $\mathcal{R}$,
to which the production cross section of the supersymmetric $t \bar t H$ process is proportional.
Since the mono-jet search is only sensitive to the stop and neutralino masses,
measuring both the mono-jet and mono-top signal rates enables us to directly probe 
the up-type higgsino components in the neutralinos through $\mathcal{R}$.
As can be seen, the sensitivity reaches up to $m_{\tilde t_1} \sim 375$ (340) (285) GeV
for $\mathcal{R} = 1$ (0.75) (0.5) at the most compressed region.
We also observe that the 2-$\sigma$ regions are almost identical between the $\tilde t_L$ and $\tilde t_R$ cases.
This means the mono-top search presented in this section works regardless of the details of the stop sector.

We finally comment on possible contributions from the $\tilde t_1$ pair production, which is not included in our calculation.
The final state of this process is two $b$-quarks from $\tilde t_1 \to b \tilde \chi_1^\pm$
and very soft fermions (possibly leptons) from $\tilde \chi_1^\pm \to f \bar f^\prime \tilde \chi_1^0$.
As shown in Fig.~\ref{fig:motivation}, the hardness of the $b$-quarks varies depending on the mass gap between $\tilde t_1$ and $\tilde \chi_1^0$,
whereas the leptons are always very soft as we fixed $\tilde \chi_1^\pm = \tilde \chi_1^0 + 3$ GeV. 
The missing energy is also tiny on average since the neutralinos are produced almost back-to-back in the transverse plane
unless they are boosted recoiling against hard ISR jets.
The efficiency of our event selection for the $\tilde t_1$ pair production
 is therefore extremely small.
This very small efficiency can however be compensated to some extent by its considerably large production rate.
We have checked numerically that the contribution from the $\tilde t_1$ pair production to our signal regions
is about 30\,\% of the $\tilde t_1 t \tilde \chi_{1(2)}^0$ contribution
in the most compressed region and rises to $\sim 70$\,\% in moderately compressed region with $\Delta m_{\tilde t_1 - \tilde \chi_1^0} \sim 50$ GeV.
This suggests that the actual sensitivity of the mono-top search is slightly better than what is shown in Fig.~\ref{fig:sensitivity},
being our results conservative.
We leave the detailed study including the $\tilde t_1$ pair production to future analyses.

\section{Probing the Stop Mixing}
\label{sec:distribution}

%
%
\begin{figure}[!t]
    \begin{center}
          \includegraphics[scale=0.4]{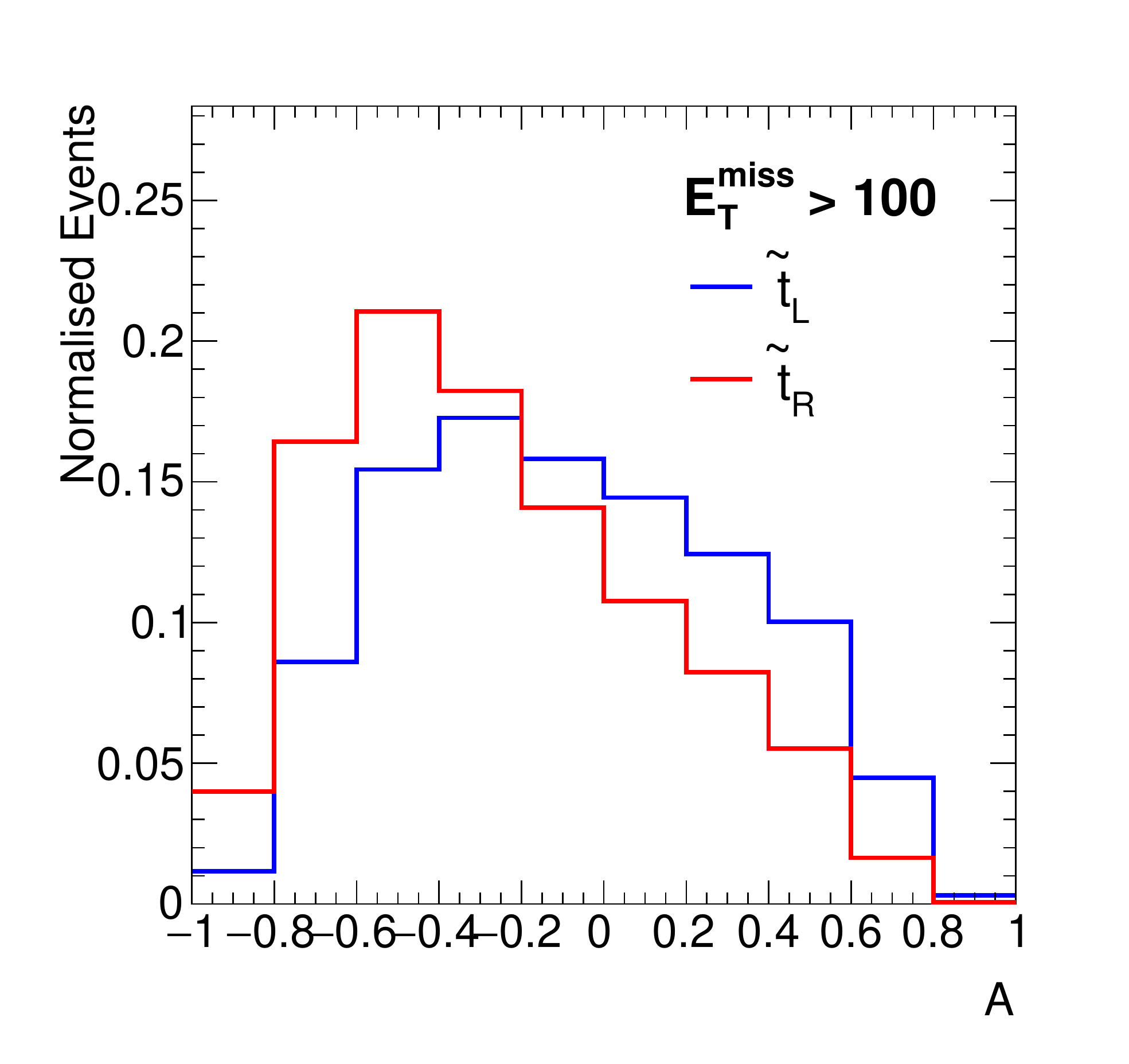}
        \hspace{-.5cm}
          \includegraphics[scale=0.4]{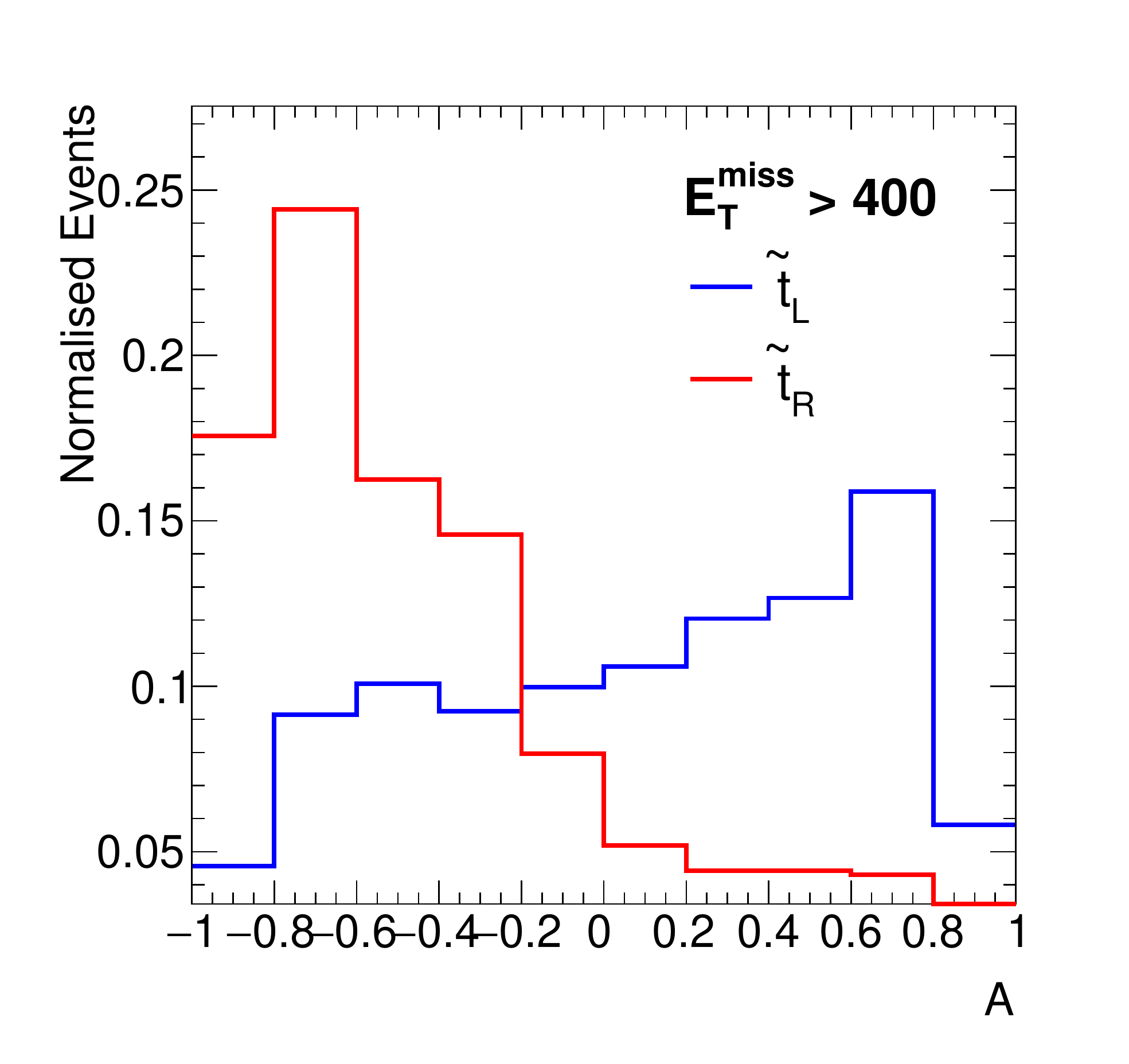}
    \caption{
    The distribution of the $p_T$ asymmetry, $\mathcal{A}$, at $(m_{\tilde t_1}, m_{\tilde \chi_1^0}) = (317, 309)$ GeV.
    The blue and red histograms correspond to the $\tilde t_1 = \tilde t_L$ and $\tilde t_1 = \tilde t_R$, respectively.
    The events satisfy the selection cuts described in Eqs.~\eqref{eq:baseline}, \eqref{eq:mblcut} and \eqref{eq:mTcut}.
    The events in the left and right plots additionally satisfy 
    $E_{T}^{\rm miss}/{\rm GeV} > 100$ and 400, respectively.
    }
    \label{fig:A}
    \end{center}
\end{figure}
%
%

We have seen that the mono-top search presented in the previous section
is insensitive to the stop mixing. The neutralino sector can be probed 
by measuring the signal rates of mono-jet and mono-top channels 
without assuming the details of the stop sector.
In this section we demonstrate, however, that 
kinematic distributions of the top-quark decay products
are sensitive to the stop sector 
and can be used to measure the stop mixing~\cite{Bhattacherjee:2012ir,Papaefstathiou:2011kd}.  

At the vicinity of the pure higgsino limit, the dominant contribution to the
stop-top-neutralino interaction comes from 
\beq
-{\cal L} \supset Y_t \Phi_{t_R} \Phi_{t^c_L} \Phi_{H_u^0} \big|_{\theta^2} 
\supset Y_t \big( \tilde t_R \bar t_L + t_R \tilde t^*_L \big) \tilde H_u^0 
\supset Y_t \big( \cos \theta_{\tilde t} \bar t_L \tilde t_1 + \sin \theta_{\tilde t} t_R \tilde t_1^* \big) 
 N_{i4} \tilde \chi_i^0 \,,
\eeq 
where $\Phi_i$ is the chiral superfield of $i$ and we have omitted the hermitian conjugate terms.
As can be seen, if $\tilde t_1$ is mostly $\tilde t_R$ ($\cos \theta_{\tilde t} \simeq 1$),
the top-quark tends to be left-handed, and vice versa for $\tilde t_L$.

The chilarity of the top-quark affects the kinematics of its decay products.
For example, the angular distribution of the decay product $f\,(= b, \ell)$ is correlated 
with the top spin direction as \cite{Jezabek:1994qs, Brandenburg:2002xr, Godbole:2010kr}
\beq
\frac{1}{\Gamma_f} \frac{d \Gamma_f}{d \cos\theta_f}=\frac{1}{2} (1 + \omega_f P_t \cos \theta_f)\, 
\eeq
in the rest frame of the top-quark, where $\theta_f$ is the angle between the decay 
product $f$ and the top spin quantization axis, 
and $P_t$ is the degree of the top polarization  
\beq
P_t \equiv \frac{N(\uparrow) - N(\downarrow)}{N(\uparrow) + N(\downarrow)}\, .
\eeq
For the top-quark in the $pp\rightarrow \tilde{t}_1t \tilde{\chi}_{1(2)}^0$ process
we obtain $P_t \simeq \cos 2 \theta_{\tilde t}$ in the pure higgsino limit.
The coefficient $\omega_f$ is given as 
$\omega_b = -0.41$ and $\omega_\ell = 1$ at tree level.

The fact that $\omega_b$ and $\omega_\ell$ have different signs means that
in the rest frame of the top-quark their momentum vectors prefer to be in the opposite direction. 
If $\tilde t_1 = \tilde t_R$ ($\cos \theta_{\tilde t} = 1$), $P_t = 1$
meaning that the boost of the top-quark is more likely to be in the direction of $\ell$
at the rest frame of the top.
In this case, the lepton gets a positive boost on average, while the $b$-quark a negative.
For $\tilde t_1 = \tilde t_L$ ($\cos \theta_{\tilde t} = 0$), 
the tendency is opposite.
To capture this feature we define the $p_T$ asymmetry, $\mathcal{A}$, as
\beq
\mathcal{A} \equiv \frac{p_T(\ell) - p_T(b)}{p_T(\ell) + p_T(b)} \,.
\eeq

We display the distribution of $\mathcal{A}$ in Fig.~\ref{fig:A} 
at $(m_{\tilde t_1}, m_{\tilde \chi_1^0}) = (317, 309)$ GeV 
for $\tilde t_1 = \tilde t_R$ (red) and $\tilde t_1 = \tilde t_L$ (blue).
We only use the events 
that pass the selection cuts Eqs.~\eqref{eq:baseline}, \eqref{eq:mblcut}, \eqref{eq:mTcut}
and $E_T^{\rm miss}/{\rm GeV} > 100$ (left panel) and 400 (right panel).
As expected, the $p_T$ asymmetry is larger (meaning that the lepton is more energetic)
for $\tilde t_L$ compared to $\tilde t_R$. The tendency is drastically enhanced if the $E_T^{\rm miss}$ threshold 
is increased from 100 to 400~GeV, because the boost of the top-quark increases. This demonstrates that  the $p_T$ 
asymmetry between the $\ell$ and $b$ is very useful variable to probe the stop mixing in the supersymmetric 
$t \bar t H$ process. 

\section{Conclusion}
\label{sec:conclusion}

In this paper, we have studied the supersymmetric  $t \bar t H$ process, i.e.~$pp \to t\, \tilde t\, \tilde h$. 
We showed that a distinctive mono-top signature arises from this channel for the Natural SUSY scenarios 
with small stop-higgisino mass differences. 
While the current searches explore this compressed stop-higgsino region 
with mono-jet channels exploiting the $\tilde t_1$ pair production associated with hard initial state radiation,
our proposed channel serves complementary bounds granting a direct probe of the the stop and neutralino sectors.

We presented a detailed search strategy to capture the supersymmetric $t \bar t H$ process  and found that a high
luminosity LHC at 13 TeV can probe the stop and higgsino sectors if $m_{\tilde t_1} \lsim 380$~GeV and 
$m_{\tilde t_1} - m_{\tilde \chi_1^0} \lsim m_W$. We observe that this sensitivity enhances for smaller mass 
differences $\Delta m_{\tilde t_1} - m_{\tilde \chi_1^0}$ and that our mono-top search works regardless of
the details of the stop sector.

Finally, we have demonstrated that the kinematic distributions of the top-quark decay products are sensitive to the
stop sector and can be used to measure the stop mixing parameter. We proposed an asymmetry variable, $\mathcal{A}$,
to access this parameter. Fortunately for our purposes, the performance of this observable dovetails nicely with the large missing energy selections required to reduce the background.

\section*{Acknowledgements}

DG and KS were supported by STFC through the IPPP grant.
The work of MT was supported by World Premier International Research Center Initiative (WPI Initiative), MEXT, Japan.

\appendix

\section{The most sensitive signal region}
\label{sec:srbest}

\begin{figure}[!t]
    \begin{center}
          \includegraphics[scale=0.43]{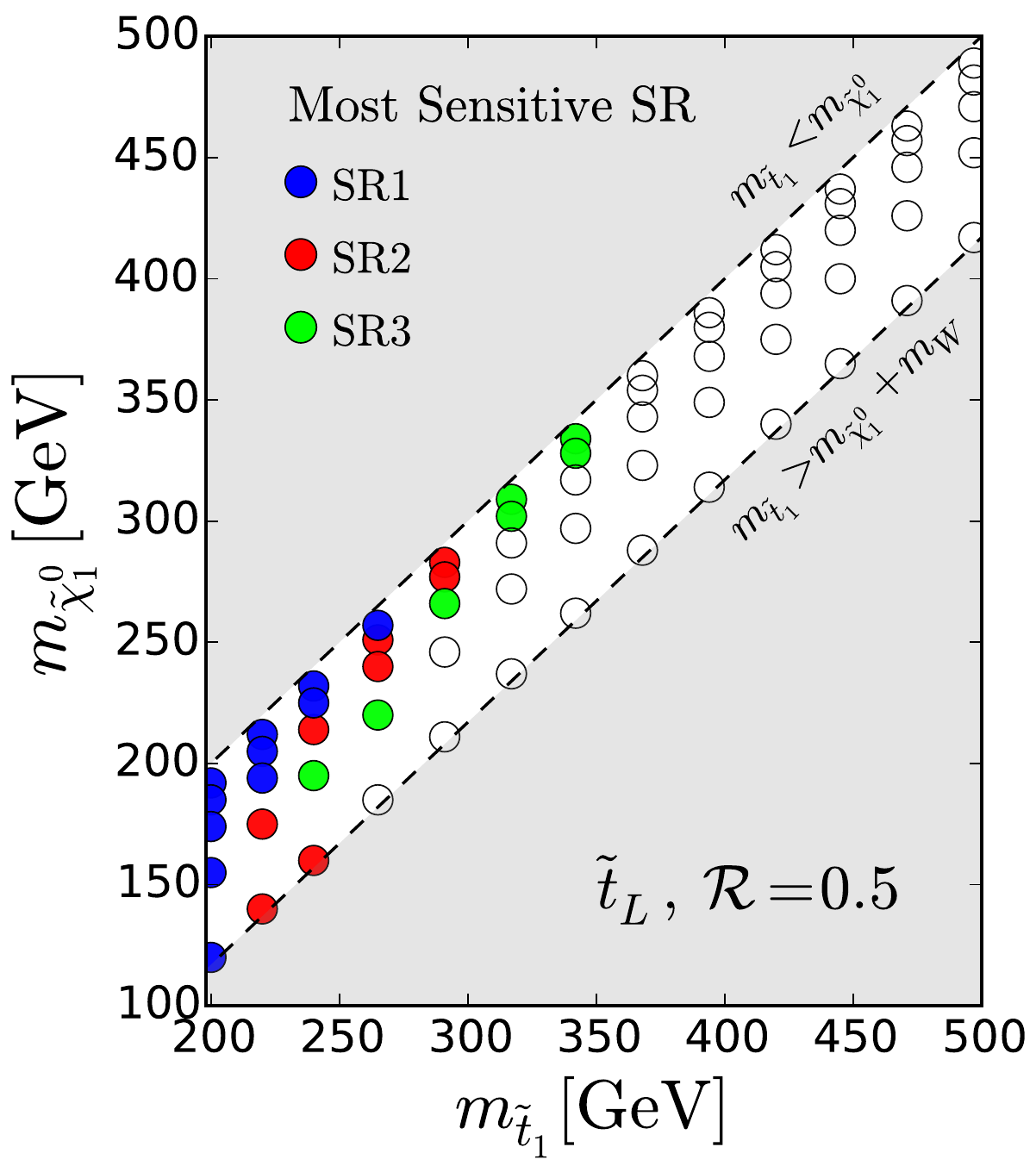}
          \includegraphics[scale=0.43]{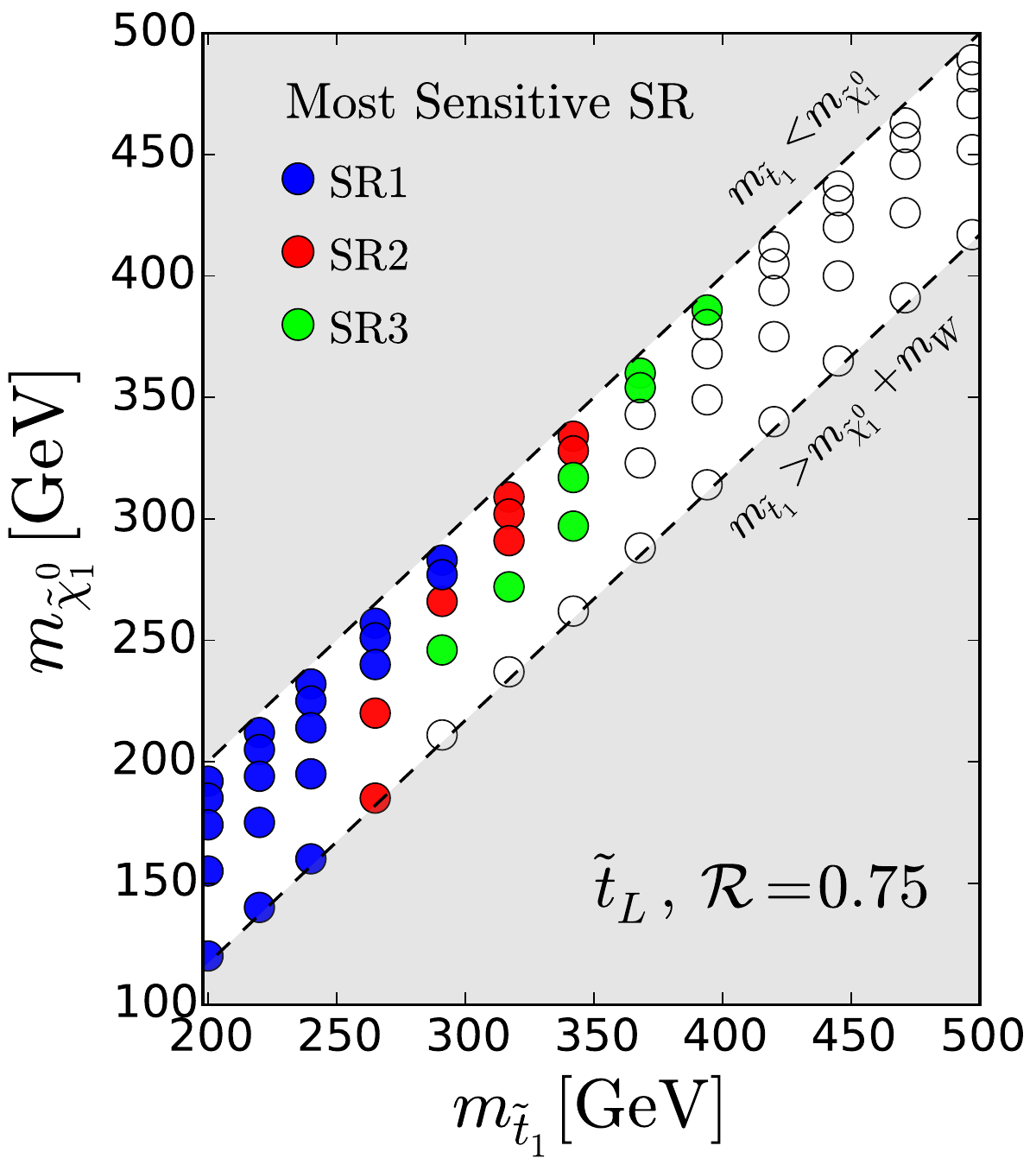}
          \includegraphics[scale=0.43]{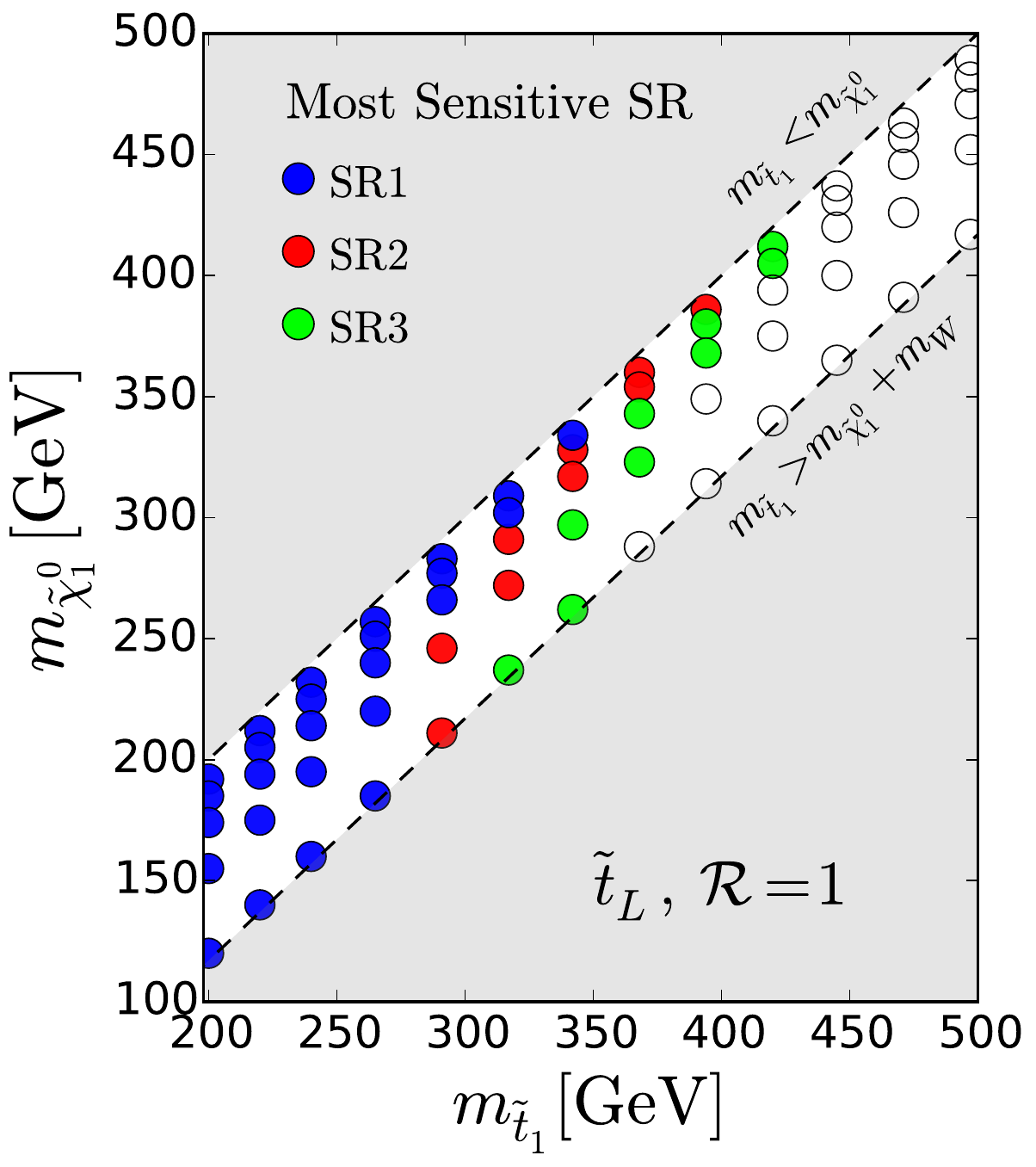}
          \\~\\
          \includegraphics[scale=0.43]{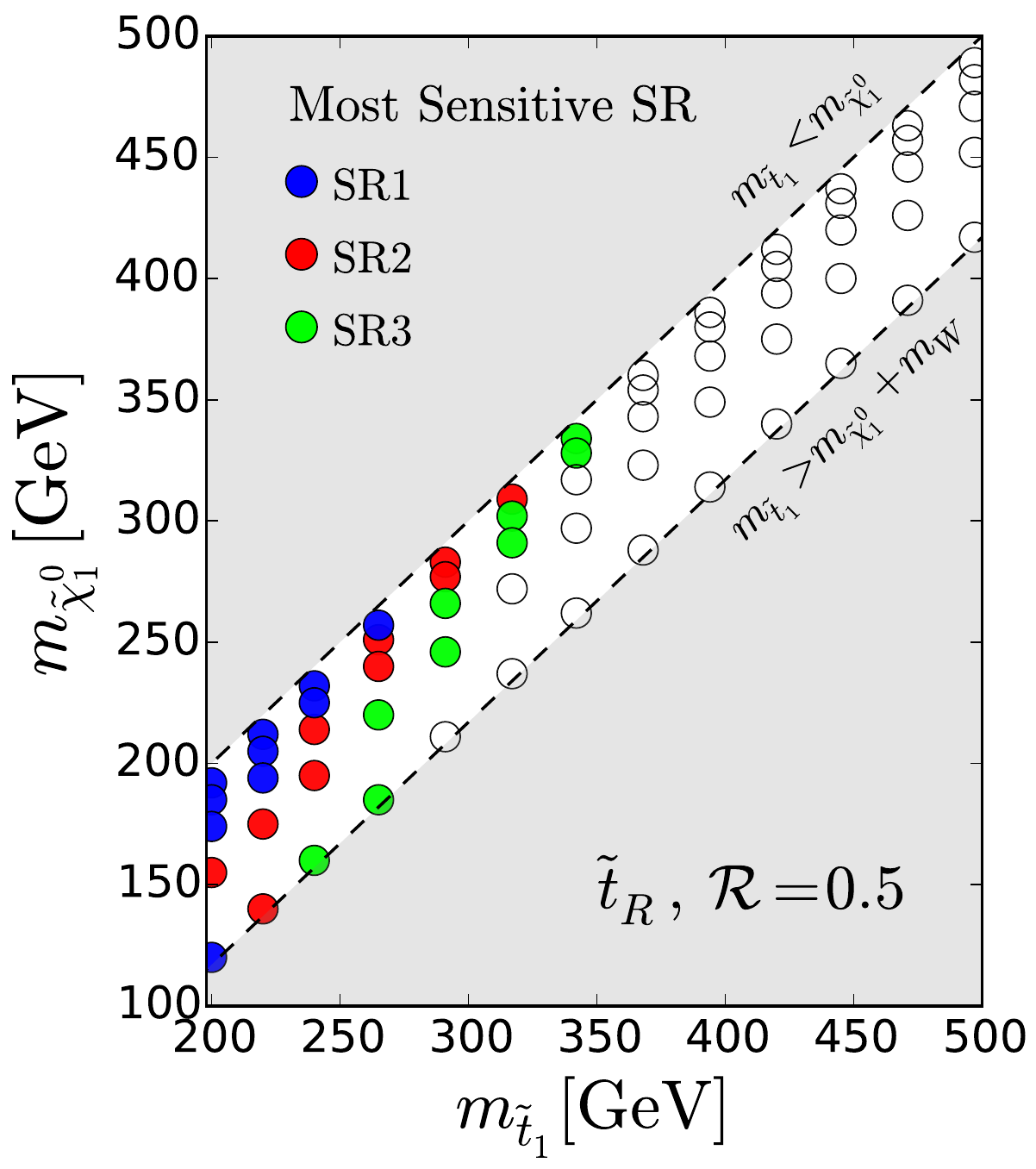}
          \includegraphics[scale=0.43]{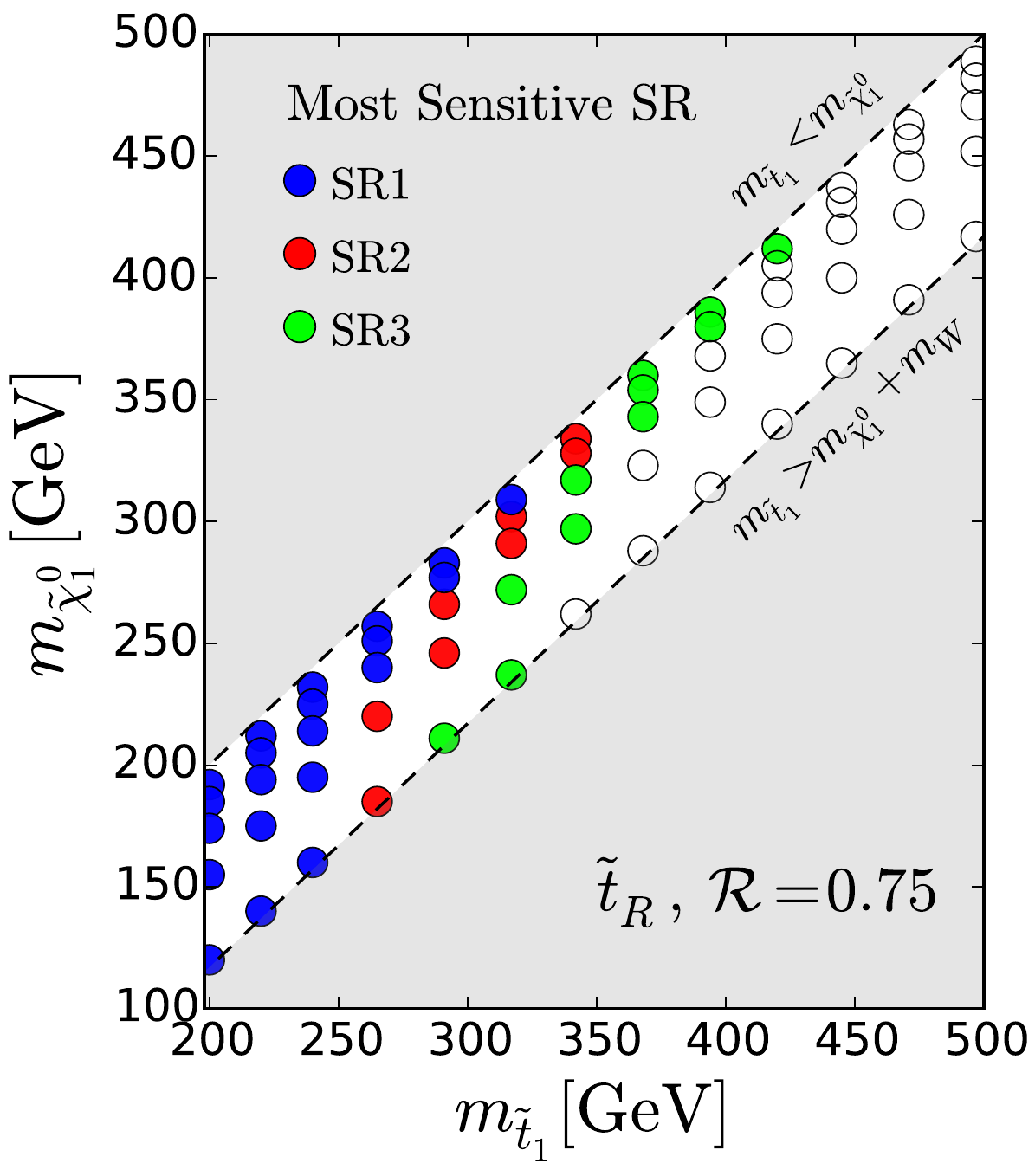}
          \includegraphics[scale=0.43]{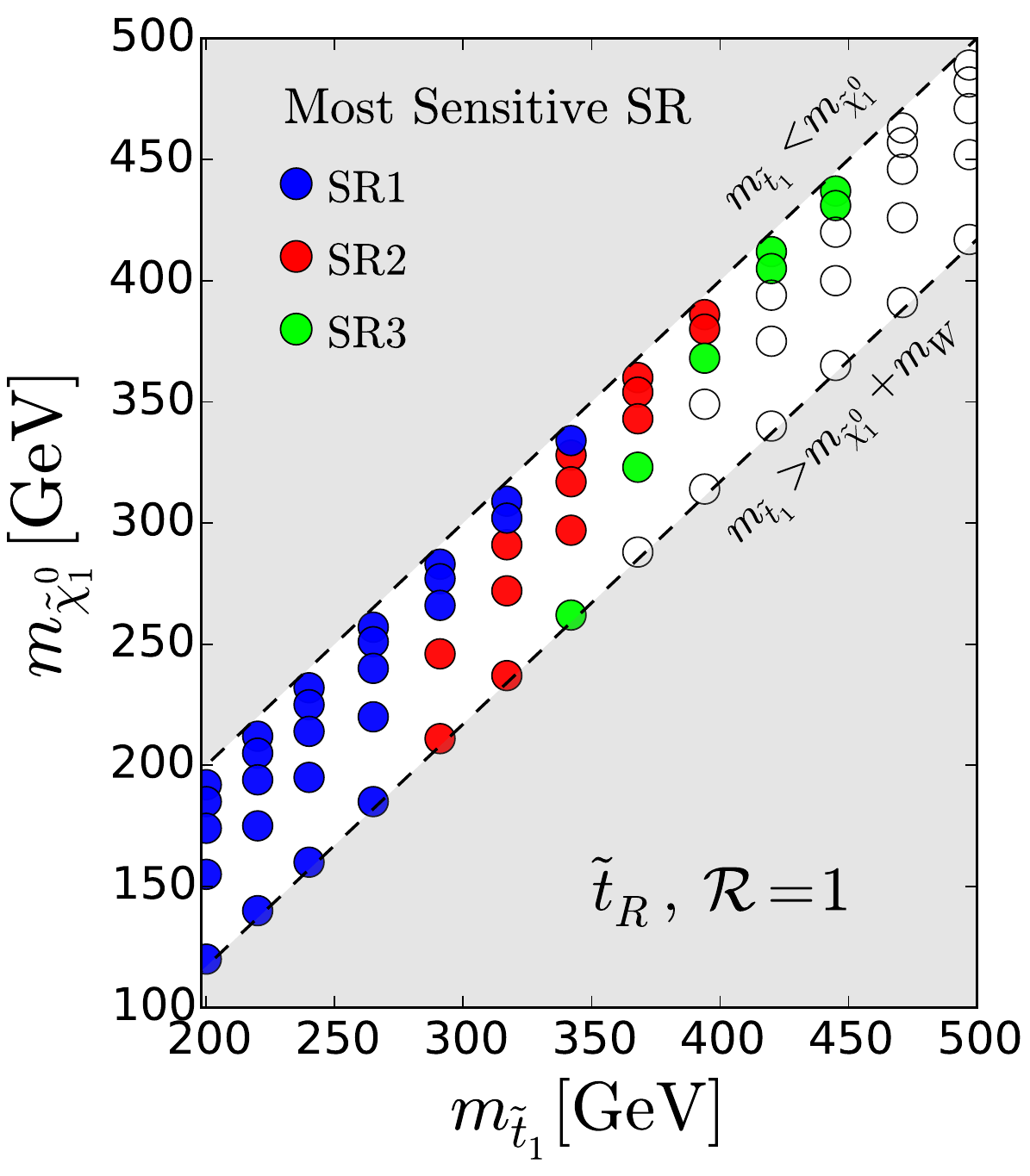}          
    \caption{
The most sensitive signal region (with the largest $S/\sqrt{B}$)
for each parameter point and for $\mathcal{R}= 0.5$ (left panel) 0.75 (centre panel) and 1 (right panel).
The top and bottom panels correspond to the $\tilde t_1 = \tilde t_L$ and $\tilde t_1 = \tilde t_R$ cases, respectively.
The empty circles represent the parameter point where none of the signal regions satisfies the consistence criteria that the
signal contribution must be greater than three and $S/B > 0.1$.    
    }
    \label{fig:srbest}
    \end{center}
\end{figure}

Fig.~\ref{fig:srbest} shows
the most sensitive signal region (with the largest $S/\sqrt{B}$)
for each parameter point and
for $\mathcal{R} = 0.5$ (left panel) 0.75 (centre panel) and 1 (right panel).
The top and bottom panel correspond to the $\tilde t_1 = \tilde t_L$ and $\tilde t_1 = \tilde t_R$ cases, respectively.
The empty circles represent the parameter points where non of the signal regions
satisfies the sanity criteria that the signal contribution must be greater than three and $S/B > 0.1$.
%
%

\newpage

\bibliographystyle{jhep}
\bibliography{draft}

\end{document}